\documentclass[
  journal=pasa,
  manuscript=research-paper,
  year=2020,
  volume=37,
]{cup-journal}

\usepackage{amsmath,amssymb}
\usepackage[nopatch]{microtype}
\usepackage{booktabs}
\usepackage{multirow}
\usepackage{array}
\newcolumntype{P}[1]{>{\raggedright\arraybackslash}p{#1}}
\newcolumntype{A}[1]{>{\raggedleft\arraybackslash}p{#1}}
\usepackage{mathtools}
\usepackage[thinc]{esdiff}

\newcommand{\bz}{$\langle B_z \rangle$}

\def\arcsec{\hbox{$^{\prime\prime}$}}

\usepackage{hyperref}
\hypersetup{colorlinks,citecolor=blue,linkcolor=blue,urlcolor=blue}

\title{Discovery of Main-sequence Radio Pulse emitters from widefield sky surveys}

\author{Barnali Das}
\affiliation{CSIRO, Space and Astronomy, P.O. Box 1130, Bentley WA 6102, Australia}
\email[Barnali Das]{Barnali.Das@csiro.au}

\author{Matt E. Shultz}
\affiliation{Department of Physics and Astronomy, University of Delaware, 217 Sharp Lab, Newark, Delaware, 19716, USA}

\author{Joshua Pritchard}
\affiliation{CSIRO, Space and Astronomy, PO Box 76, Epping, NSW 1710, Australia}

\author{Kovi Rose}
\affiliation{Sydney Institute for Astronomy, School of Physics, The University of Sydney, New South Wales 2006, Australia}
\alsoaffiliation{CSIRO, Space and Astronomy, PO Box 76, Epping, NSW 1710, Australia}

\author{Laura~N. Driessen}
\affiliation{Sydney Institute for Astronomy, School of Physics, The University of Sydney, New South Wales 2006, Australia}

\author{Yuanming Wang}
\affiliation{Centre for Astrophysics and Supercomputing, Swinburne University of
Technology, Hawthorn, Victoria, 3122, Australia}

\author{Andrew Zic}
\affiliation{CSIRO, Space and Astronomy, PO Box 76, Epping, NSW 1710, Australia}

\author{Tara Murphy}
\affiliation{Sydney Institute for Astronomy, School of Physics, The University of Sydney, New South Wales 2006, Australia}

\author{Gregory Sivakoff}
\affiliation{Department of Physics, University of Alberta, CCIS 4-181, Edmonton AB T6G 2E1, Canada}


\begin{document}

\begin{abstract}
Magnetic AB stars are known to produce periodic radio pulses by the electron cyclotron maser emission (ECME) mechanism. Only 19 such stars, known as `Main-sequence Radio Pulse emitters' (MRPs) are currently known. The majority of MRPs have been discovered through targeted observation campaigns that involve carefully selecting a sample of stars that are \textit{likely} to produce ECME, and which can be detected by a given telescope within reasonable amount of time. These selection criteria inadvertently introduce bias in the resulting sample of MRPs, which affects subsequent investigation of the relation between ECME properties and stellar magnetospheric parameters. The alternative is to use all-sky surveys. Until now, MRP candidates obtained from surveys were identified based on their high circular polarisation ($\gtrsim 30\%$). In this paper, we introduce a complementary strategy, which does not require polarisation information. Using multi-epoch data from the Australian SKA Pathfinder (ASKAP) telescope, we identify four MRP candidates based on the variability in the total intensity light curves. Follow-up observations with the Australia Telescope Compact Array (ATCA) confirm three of them to be MRPs, thereby demonstrating the effectiveness of our strategy. With the expanded sample, we find that ECME is affected by temperature and the magnetic field strength, consistent with past results. There is, however, a degeneracy regarding how the two parameters govern the ECME luminosity for magnetic A and late-B stars (effective temperature $\lesssim 16$ kK). The current sample is also inadequate to investigate the role of stellar rotation, which has been shown to play a key role in driving incoherent radio emission. 
\end{abstract}

\section{Introduction}\label{sec:intro}
Magnetic hot stars are stars of spectral types O, B and A, characterised by the presence of large-scale, kG-strength, highly stable surface magnetic fields. The magnetic field can usually be approximated as a dipole inclined to the stellar rotation axis \citep[e.g.][]{shultz2018}. The magnetic field traps plasma from the stellar wind to form a co-rotating magnetosphere. \citet{drake1987} first discovered non-thermal radio emission from a sample of magnetic hot stars, thereby proving the existence of a particle acceleration mechanism in the stellar magnetosphere that provides the relativistic particles. These electrons, while gyrating in the magnetic field, emit gyrosynchrotron emission. \citet{linsky1992} proposed that the incoherent radio luminosity correlates with the stellar magnetic field and effective temperature. This inference was drawn using a sample of 16 stars. The scenario changed completely when \citet{leto2021} and \citet{shultz2022} investigated the relation between incoherent radio luminosity and stellar parameters using samples of 30 and 50 stars respectively.
Both studies showed that incoherent radio luminosity is determined by three stellar parameters: magnetic field strength, stellar radius, and rotation period. These results motivated the development of a new scenario of radio production in large-scale stellar magnetospheres, which is the centrifugal breakout \citep[CBO;][]{townsend2005} driven scenario \citep{owocki2022}. According to this scenario, electrons are accelerated via magnetic reconnection, triggered by CBOs, which are small spatial-scale explosions caused by temporary disruption of magnetic field lines by excess centrifugal force acting on co-rotating magnetically trapped plasma.

The evolution in the understanding of gyrosynchrotron emission from magnetic hot stars demonstrates the importance of having a large sample size to develop robust theories. However, in addition to gyrosynchrotron emission, a subset of magnetic hot stars are known to produce pulses of coherent emission driven by the electron cyclotron maser emission (ECME) mechanism \citep[e.g.][]{trigilio2000,das2022}. To develop a complete understanding of non-thermal radio emission in hot star magnetospheres, it is equally important to investigate the coherent component.
\citet{trigilio2000} first discovered coherent radio emission produced by electron cyclotron maser emission (ECME) from the magnetic hot star CU\,Vir. 
Such stars are now referred as Main-sequence Radio Pulse emitters \citep[`MRPs',][]{das2021}.

ECME is intrinsically a narrow-band emission with the frequency being proportional to the magnetic field strength at the emission sites. As a result, higher frequencies are produced closer to the star. For mildly relativistic electrons, the emission is highly directed at almost $90^\circ$ to the local magnetic field direction at the site where the suitable conditions for powering the maser are realised \citep{melrose1982}. Consequently, ECME is observed as pulses at rotational phases when the line-of-sight crosses the emission beam originating from the envelope of the elementary beams. 

Prior to 2018, only two MRPs were known \citep{trigilio2000, chandra2015,das2018}. Both discoveries were serendipitous. Since then, 17 more MRPs have been reported \citep{lenc2018,leto2019,das2019a,das2019b,leto2020,leto2020b,das2022,das2022c,biswas2025}, 13 of which were from a carefully designed targeted campaign \citep{das2019a,das2019b,das2022,das2022c}.
Thus, we currently know of 19 MRPs.

Unlike the incoherent component of non-thermal radio emission from magnetic hot stars, the driving mechanism of ECME production and the role of different stellar parameters remain unclear. \citet{leto2021} suggested that the engine behind the coherent emission could be different from that behind the incoherent gyrosynchrotron emission. \citet{das2022c} compared the spectral incoherent and coherent radio luminosity for the MRPs and found that the two exhibit a strong positive correlation for stars with effective temperatures $T_\mathrm{eff}\lesssim 19$ kK. This prompted them to examine the correlation between the coherent radio luminosity and the same stellar parameters that were found to be responsible for gyrosynchrotron emission. Although they found a clear positive correlation with magnetic field strength \citep[also obtained by][]{das2022}, the role of stellar radius and rotation period remained inconclusive with their sample size.

In order to perform a robust study of the relationship between ECME and stellar magnetospheric parameters, expanding the sample spanning a wider parameter-space is necessary. However, to date samples of MRPs have been dominated by discoveries made from targeted observations that introduce selection bias. This issue can be overcome by using all-sky surveys to find MRP candidates, which can then be confirmed by dedicated observations. 
Among the 19 known MRPs, two were discovered as MRP candidates based on their high circular polarisation from widefield survey data \citep{lenc2018,pritchard2021}. This strategy of searching for highly circularly polarised emission is highly effective at finding coherent radio emitters since incoherent emission is only weakly circularly polarised. Here we demonstrate a complementary strategy that relies on multi-epoch data to find MRP candidates, and does not require polarisation information. By applying this strategy to data acquired by the Australian SKA Pathfinder (ASKAP) radio telescope, we identify four MRP candidates, which were followed-up with the Australia Telescope Compact Array (ATCA). We confirm three of them to be MRPs.

This paper is structured as follows. In the next section, we describe the limitations of targeted observations and how all-sky surveys address those problems (\S\ref{sec:MRP_detection_complementary_strategy}). This is followed by a description of our targets (\S\ref{sec:target}), and ASKAP and ATCA observations and data analysis (\S\ref{sec:observation}). We present the results and discussion in sections \S\ref{sec:results} and \S\ref{sec:discussion} respectively, and conclude in \S\ref{sec:conclusion}.

\section{Need for a complementary strategy to discover MRPs}\label{sec:MRP_detection_complementary_strategy}
Targeted observations have proved to be a successful means of discovering MRPs \citep[e.g.][]{das2022}. The strategy is to observe a sample of magnetic hot stars over a rotational phase range during which ECME pulses are expected to arrive. However, this strategy introduces biases into the sample of confirmed MRPs in the following ways:
\begin{enumerate}
    \item Selection criteria for constructing the sample: It is desirable to include stars that are likely to produce ECME. This involves making assumptions about which stars are likely to produce the emission and which are not. For example, no stars with polar magnetic field strength $\lesssim 1.5$\,kG have been included in previous targeted campaigns \citep{das2022c,das2022}. Apart from that, no star with a rotation period longer than a few days is included in targeted campaigns as the amount of time required to cover a given range of rotational phase range increases with increasing rotation period.
    \item Choice of rotational phase range: Due to limited telescope time, targeted campaigns cannot observe each star for its full rotation cycle. One has to predict the arrival phases of the radio pulses. This is done using the ephemeris, stellar magnetospheric geometry, and a model for emission beaming geometry. For magnetic hot stars, the `tangent plane beaming model' \citep{trigilio2011} is used to predict ECME arrival phases. It predicts that ECME is visible around the rotational phases at which the stellar longitudinal magnetic field $\langle B_\mathrm{z} \rangle$ is zero (called magnetic nulls). Thus, observations are scheduled around the magnetic null phases. Because of this constraint, only stars with well-constrained magneto-rotational properties can be included.
\end{enumerate}

\begin{figure*}
\centering
    \includegraphics[width=0.48\textwidth]{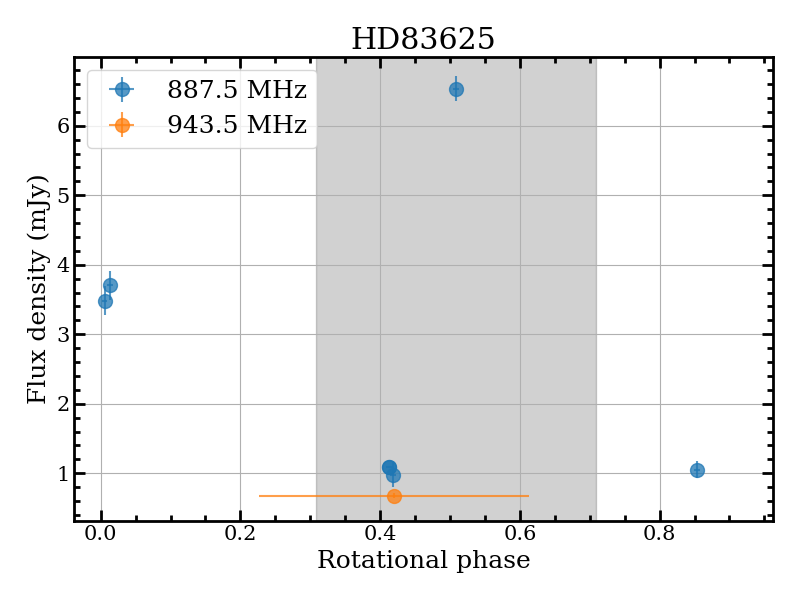}
    \includegraphics[width=0.48\textwidth]{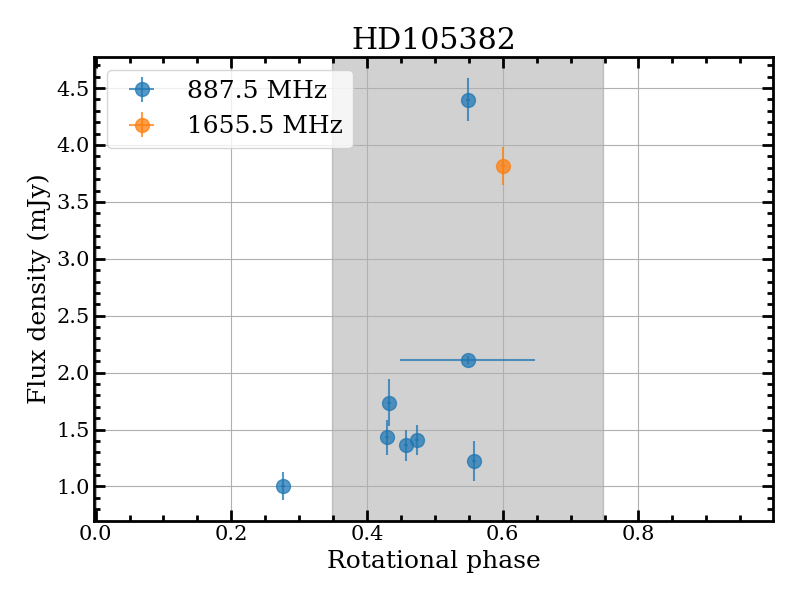}
    \includegraphics[width=0.48\textwidth]{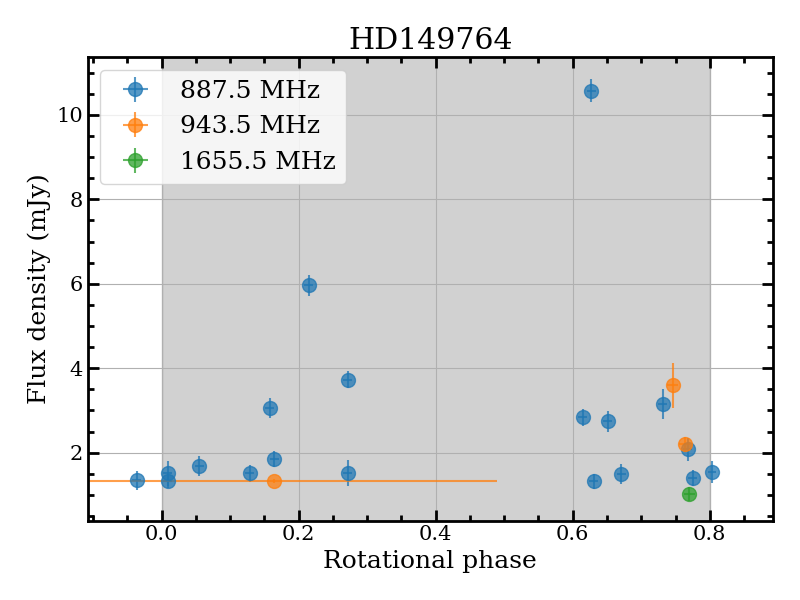}
    \includegraphics[width=0.48\textwidth]{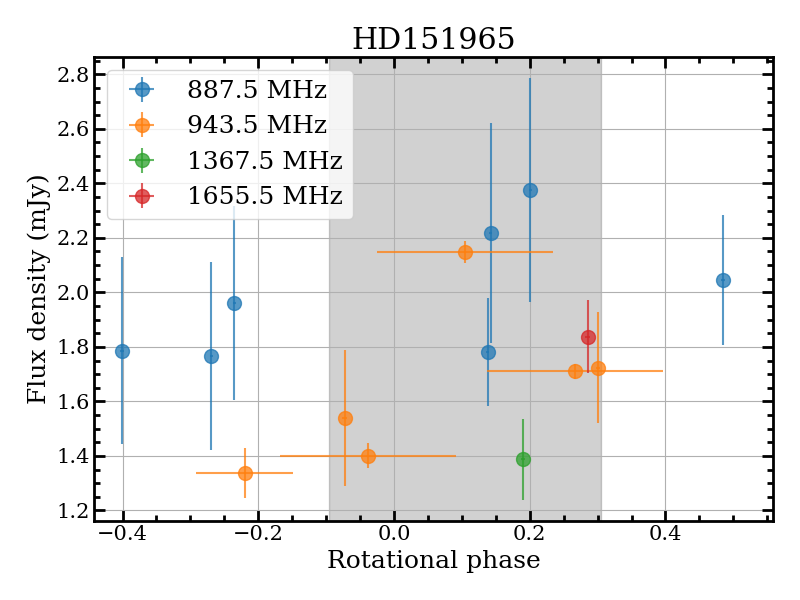}
    \caption{The light curves of the four targets that were identified as MRP candidates based on the variation of their flux densities (total intensity). The errorbars associated with the rotational phases indicate the integration times corresponding to the flux density measurements. All data were acquired with the ASKAP. The data are phased using the ephemerides given in Table \ref{table:ephemeris}. The rotational phase window chosen to cover with the ATCA are marked with gray shaded regions. Note that the actual coverages of the rotational phases are slightly different from those shown here.}
    \label{fig:askap_lc}
\end{figure*}

To construct a less biased sample, we need to use widefield radio surveys. The principal strategy is to search for magnetic hot stars whose radio properties are uncharacteristic of incoherent gyrosynchrotron radio emission. There are two such properties of ECME that can be used. The first is the
high circular polarisation. ECME can be up to $100$\% circularly polarised and the sense of polarisation is opposite for opposite magnetic field directions. Gyrosynchrotron emission is only mildly circularly polarised ($\lesssim 20\%$) and at $\lesssim 1$ GHz bands, the circular polarisation is nearly insignificant \citep[e.g.][]{leto2017,dulk1985}. Thus, high circular polarisation is a strong indicator of ECME. This strategy is widely used not only for magnetic hot stars, but also to identify coherent radio emission from ultracool dwarfs and brown dwarfs \citep[e.g.][]{vedantham2022,rose2023}.

The second property is high variability. With the growing sample of MRPs, it has emerged that the characteristics of radio pulses from MRPs show diverse behaviour, and high circular polarisation is not a universal pulse property \citep[e.g.][]{das2022}. But the one property that is found to be valid in all cases is the much shorter timescale of the variation of flux density as compared to that of the incoherent radio emission (which modulates with the stellar rotation period). This property is associated with the high directivity of ECME. \citet{das2022} introduced the `minimum flux-density gradient condition': 
\begin{align}
    \Delta\phi_\mathrm{rot}<\frac{1}{2\pi}\approx0.16, \label{eq:variability_condition}
\end{align}
where $\Delta \phi_\mathrm{rot}$ is the rotational phase range over which the pulse rises from its basal to peak flux density, as a necessary condition for distinguishing MRP behaviour. 
Note that the above condition is derived assuming that incoherent flux density modulates as $\sin^2(2\pi\phi_\mathrm{rot})$, where $\phi_\mathrm{rot}$ is the stellar rotational phase \citep{das2022}.


The advantage of using the criterion of high variability over that of high circular polarisation is that one does not need polarisation information to find MRP candidates. The disadvantage is that one needs a light curve to identify MRP candidates, which is not a typical data-product of wide-field surveys. This is where 
the `Variables and Slow Transients' \citep[VAST;][]{murphy2013} survey, being carried out with the ASKAP, offers a unique advantage. 
VAST observes the Galactic regions with a fortnightly cadence and the extragalactic regions every two months. The integration time per epoch is $12$ minutes, which is small enough for magnetic hot stars that have rotation periods from half a day to decades \citep[][Shultz et al. in prep.]{shultz2018,shultz2019c,shultz2019b,shultz2020}, so as not to average out any flux density enhancement occurring over a small rotational phase range.
Thus, we obtain light curves for all magnetic hot stars covered by the VAST survey. These are further supplemented with data from other ASKAP surveys such as the Rapid ASKAP Continuum Survey \citep[RACS;][integration time of 15 minutes]{mcconnell2020} and the Evolutionary Map of the Universe \citep[EMU;][integration time of 10 hours]{emu2011} which observes at $943.5$\,MHz.
RACS has three components based on the central frequency: RACS-Low \citep[$887.5$\,MHz,][]{hale2021}, RACS-Mid \citep[$1367.5$\,MHz,][]{duchesne2023,duchesne2024} and RACS-High \citep[$1655.5$\,MHz,][]{duchesne2025}. These light curves can be used to search for MRP candidates.

Note that ASKAP also provides polarisation information. 
In this paper, we present MRP candidates discovered purely using our variability criterion and their follow-up observation with the ATCA.

\begin{table}
\begin{tabular}{lll}
\toprule
\headrow Star & $\mathrm{HJD_0}$ (days) & $P_\mathrm{rot}$ (days) \\
\midrule 
HD\,83625 & 2460015.42769 & 1.0784747(1)\\
HD\,105382 & 2460041.62838 & 1.2950709(2)\\
HD\,149764 & 2460097.82080 & 0.63934468(7)\\
HD\,151965 & 2460098.35344 & 1.60866(3)\\
\midrule
\caption{The ephemerides used to calculate the rotational phases of the star (\S\ref{sec:rotation_period}). 
}
\label{table:ephemeris}
\end{tabular}
\end{table}

\section{Our targets}\label{sec:target}
We identified radio emitting magnetic hot stars using the cross-match methods and catalogues described in \citet{driessen2024}. We cross-matched a catalogue of magnetic hot stars (Shultz et al. in prep) with the combined ASKAP catalogues described in \citet{driessen2024} and identified 37 radio emitting magnetic hot stars. Following the methods in \citet{driessen2024}, we used cross-match radii with at least 98\% reliability. Due to the low sky density of magnetic hot stars, this cross-match radius was 4\arcsec\ for all cross-matches. See \citet{driessen2024} for a full explanation of the cross-match method and reliability calculations.

We identified four candidate MRPs from the sample of 37 radio emitting magnetic hot stars by examining their light curves. The light curves were extracted using the publicly available data in the CSIRO ASKAP Science Data Archive (CASDA\footnote{\url{https://research.csiro.au/casda/}}). Of the 37 radio emitting magnetic hot stars, 
and
20 had detections at more than two epochs. As we used all publicly available ASKAP data, the light curves include observations with integration times from 12 minutes to 10 hours and central frequencies ranging from 887.5 MHz to 1655.5 MHz. 
Since coherent radio emission is known to exhibit significant variation with frequencies \citep[e.g.][]{das2021}, we construct separate light curves for each of these four frequencies (where possible), which are 887.5 MHz, 943.5 MHz, 1367.5 MHz and 1655.5 MHz.

In order to check for the validity of Equation \ref{eq:variability_condition}, we converted the time axis to rotational phase axis for the stars with rotation period information. 
These light curves are visually inspected to identify MRP candidates on the basis of variability at any of the four frequencies. This process resulted into four MRP candidates: HD\,83625, HD\,149764, HD\,105382, and HD\,151965, which we followed up with the ATCA. 
Note that HD\,105382 was previously detected in a circular polarisation survey with a fractional circular polarisation of 60\% \citep{pritchard2021}, further suggesting that the star is an MRP.

For the four chosen targets, the rotation periods are refined from their literature values using optical light curves provided by the Transiting Exoplanet Survey Satellite, (see \S\ref{sec:rotation_period}). The newly obtained rotation periods and the reference epochs are listed in Table \ref{table:ephemeris}. These ephemerides will be used throughout the paper.

In Figure \ref{fig:askap_lc}, we show the light curves obtained from ASKAP data. The reference epochs and rotation periods used to calculate the stellar rotational phases are listed in Table \ref{table:ephemeris}. The majority of the data are acquired at $887.5$\,MHz. The stars HD\,83625, HD\,105382, and HD\,149764 exhibit strong variation in flux densities at $887.5$\,MHz and comfortably satisfy the minimum flux density gradient condition (\S\ref{sec:MRP_detection_complementary_strategy}). HD\,151965 is the only candidate that exhibits variability consistent with Equation \ref{eq:variability_condition} at 943.5 MHz.

Among these four stars, HD\,83625 and HD\,149764 respectively have only one and two longitudinal magnetic field measurements \citep{bagnulo2015}, and thus their \bz~curves are not known. For HD\,149764, the sign of \bz~for the two measurements are opposite showing that the \bz~curve exhibits magnetic nulls, though locating the precise rotation phase at which they occur is not possible. For HD\,83625, it is not known a priori whether or not its \bz~curve exhibits magnetic nulls. In other words, stars like these are unlikely to be investigated under targeted campaigns. Magnetic data of HD\,151965 were reported by \citet{bohlender1993} and those for HD\,105382 were reported by \citet{briquet2007}. The \bz~curve for HD\,105382 exhibits magnetic nulls, but that for HD\,151965 does not.

\section{Observations} \label{sec:observation}
We observed our four MRP candidates with the ATCA at $1$--$3$\,GHz (project code: C3608). Barring HD\,149764, the stars were observed over a rotational phase range of $\pm0.2$ cycles centred around the rotational phase corresponding to the maximum flux density observed with the ASKAP. HD\,149764, shows clear indication of multiple flux density enhancements in the light curve obtained with ASKAP data (Figure \ref{fig:askap_lc}) and also has the lowest rotation period among the four stars \citep[$\approx 0.64$ days,][]{shultz2022}. This star was, hence, observed over a rotational phase window of $0.8$ cycles (twice that of the other stars). For all observations, the maximum baseline was 6 km. We used PKS 1934--638 as the primary calibrator. The secondary calibrators used are 0939--608 (HD\,83625), 1234--500 (HD\,105382) and 1714--397 (HD\,149764 and HD\,151965). The properties of these sources can be found in the ATNF calibrator database\footnote{https://www.narrabri.atnf.csiro.au/calibrators/calibrator\_database.html}.

Although the raw data span the frequency range of $1076$--$3124$ MHz, a significant portion of the lower frequency edge of the band need to be flagged due to radio frequency interference (RFI). After flagging, the effective band spans the frequency range of $1293$--$3023$ MHz. All observations were flagged and calibrated using a standard continuum data reduction routines with \textsc{miriad}. The calibrated full Stokes data were then imaged and self-calibrated using \textsc{wsclean} \citep{offringa2014} and \textsc{CASA} \citep{mcmullin2007}. We then extracted dynamic spectra in all Stokes parameters using \textsc{DStools}\footnote{\url{https://github.com/askap-vast/dstools}} (Pritchard et al., in prep.) by subtracting a target-masked field model from the self-calibrated visibilities to remove contamination from field sources, and then vector-averaging the subtracted visibilities over all baselines. We further time- and frequency-averaged the dynamic spectra to optimise signal-to-noise and time/frequency resolution of pulse features.

We use the IAU/IEEE convention for defining right and left circular polarisations (RCP and LCP respectively) and for defining the sign of Stokes V.  Stokes I and V are related to RCP and LCP by the following equations:
\begin{align*}
    \mathrm{I}&=\frac{\mathrm{RCP}+\mathrm{LCP}}{2}\\
    \mathrm{V}&=\frac{\mathrm{RCP}-\mathrm{LCP}}{2}
\end{align*}
Thus, positive Stokes V indicates an enhancement in RCP and vice-versa.

\section{Results}\label{sec:results}
We detect strong flux density enhancements from three of our targets, confirming them as MRPs. The only star that we could not confirm as MRP is HD\,151965. Below we describe the results obtained for each target in greater detail.

\subsection{HD\,83625}\label{subsec:hd83625}

\begin{figure}
    \centering
    \includegraphics[width=0.95\textwidth]{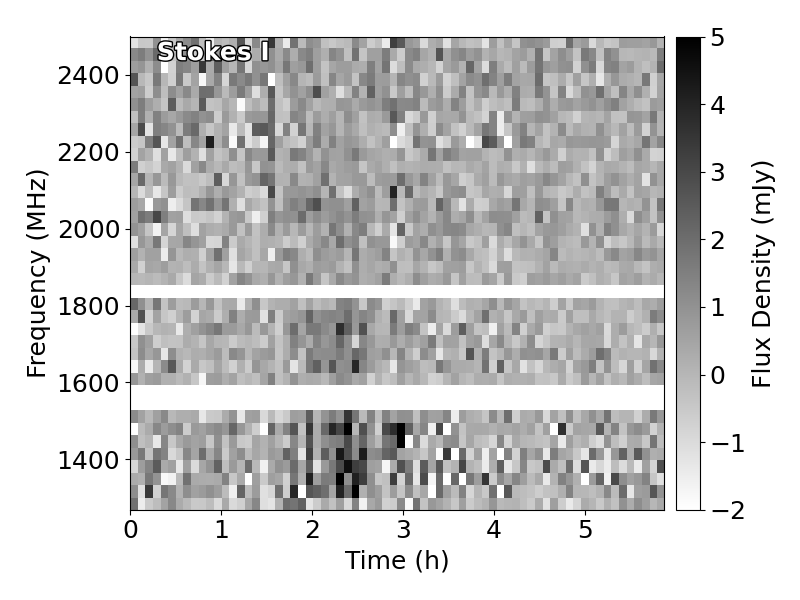}
    \includegraphics[width=0.95\textwidth]{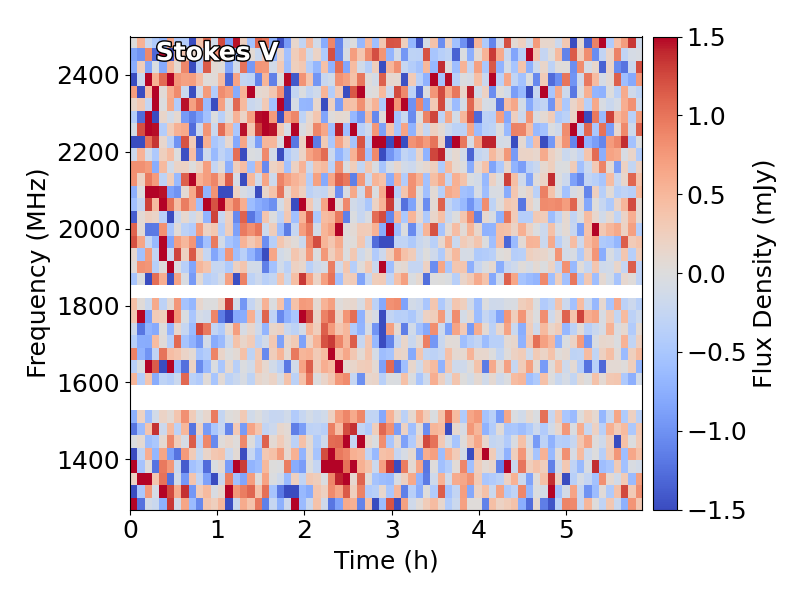}    
    \caption{The dynamic spectra of HD\,83625 in Stokes I and V up to a frequency of $2500$ MHz,
    averaged with $5$ minute time resolution and $32$\,MHz frequency resolution.
    The horizontal gaps mark the flagged channels.}
    \label{fig:hd83625_IV_DS}
\end{figure}

\begin{figure}
    \centering
    \textbf{\hspace{1cm}HD\,83625}\par\medskip
    \includegraphics[trim={0 0 0 0.8cm},clip,width=0.95\textwidth]{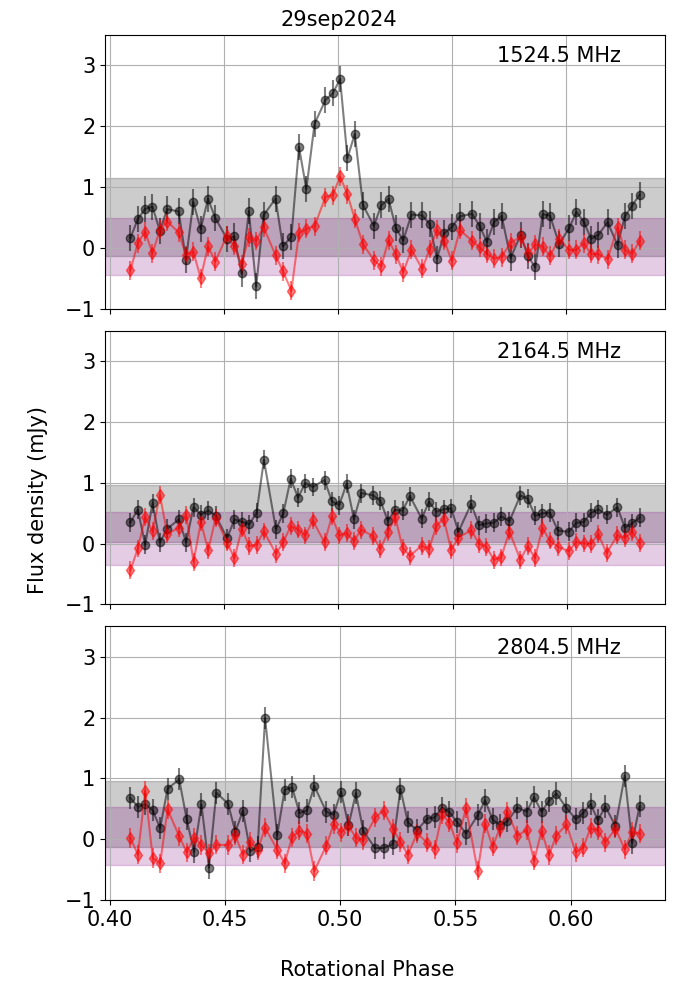}
    \caption{The Stokes I (black) and V (red) light curves of HD\,83625 extracted from the visibility domain in three frequency subbands. The integration time for each measurement is 5 minutes. The shaded regions indicate the 3$\sigma$ variation about the median flux density away from the phases of enhancement.
    }
    \label{fig:hd83625_IV_lc}
\end{figure}

HD\,83625 was observed on 2024--09--29. Its dynamic spectra in Stokes I and V are shown in Figure \ref{fig:hd83625_IV_DS}. The total intensity plot shows an enhancement in flux density below $1800$\,MHz, with a weak signature in the Stokes V dynamic spectrum. 
We divided the full band into three sub-bands, centred at $1524.5$\,MHz, $2164.5$\,MHz and $2804.5$\,MHz, with a bandwidth of 640 MHz each, and extracted light curves by averaging over each of these sub-bands. The result is shown in Figure \ref{fig:hd83625_IV_lc}.
There is a clear flux density enhancement at the lowest frequency bin ($1524.5$\,MHz) observed at phase of $\approx 0.09$, which is very close to the rotational phase of enhancement (0.1) observed at 887.5 MHz with ASKAP (see Figure \ref{fig:askap_lc}). This confirms that the star is an MRP.
The corresponding Stokes V light curve shows a weak, but prominent enhancement around 0.09 phase. The maximum observed flux density at $1524.5$\,MHz ($\approx2$\,mJy) is significantly smaller than that observed at $887.5$\,MHz ($\approx6$\,mJy). This is consistent with the fact that the pulse exhibits a cut-off within our observed frequency range (see Figure \ref{fig:hd83625_IV_DS}).

\begin{figure}
    \centering
    \includegraphics[width=0.98\textwidth]{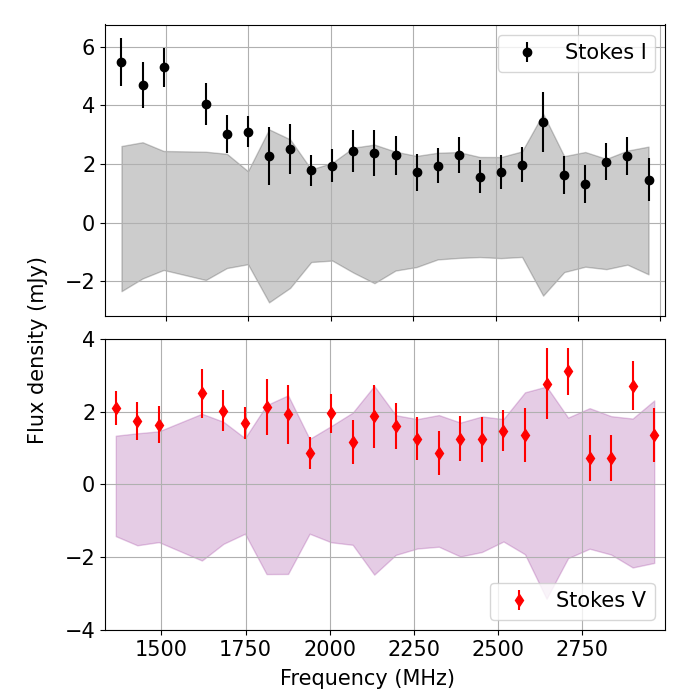}
    \caption{The peak flux density spectrum for the pulse observed from HD\,83625. The shaded regions indicate 3$\sigma$ variations about the basal flux density spectrum.}
    \label{fig:hd83625_I_spectrum}
\end{figure}

To locate the upper cut-off frequency more precisely, we extracted the peak flux density spectra for total intensity and Stokes V. The maximum observed circular polarisation is $38\pm10\%$ at $1364.5$\,MHz (with an integration time of 5 minutes).
We define the upper cut-off frequency as the lowest frequency at which the peak flux density becomes consistent with the basal flux density $\pm 3\sigma$. We find that the Stokes V spectra indicates a cut-off at $1428\pm32$\,MHz (see Figure \ref{fig:hd83625_I_spectrum}). This is, however, clearly an underestimate of the upper cut-off frequency since the corresponding flux density in total intensity is significantly higher than the respective basal flux density despite the somewhat larger noise in the Stokes I spectrum. This is not a surprise as the observed circular polarisation of the ECME pulses is expected to go to zero near the upper cut-off frequency due to geometric effects \citep{leto2016,das2020a}. We hence conclude that the true upper cut-off frequency ($1812\pm 32$\, MHz) is indicated by the Stokes I peak flux density spectrum (Figure \ref{fig:hd83625_I_spectrum}). This is significantly smaller than the gyrofrequency\footnote{Electron gyrofrequency $\nu_\mathrm{B}\approx 2.8 B$, where $B$ is in gauss and $\nu_\mathrm{B}$ is in MHz.} corresponding to the only available \bz~measurement for the star \citep[$3.5$ GHz, corresponding to $-1245\pm 77$\,G,][]{bagnulo2015}, showing that this MRP exhibits a premature upper cut-off.


\subsection{HD\,105382}\label{subsec:hd105382}
\begin{figure}
    \centering
    \includegraphics[width=\columnwidth]{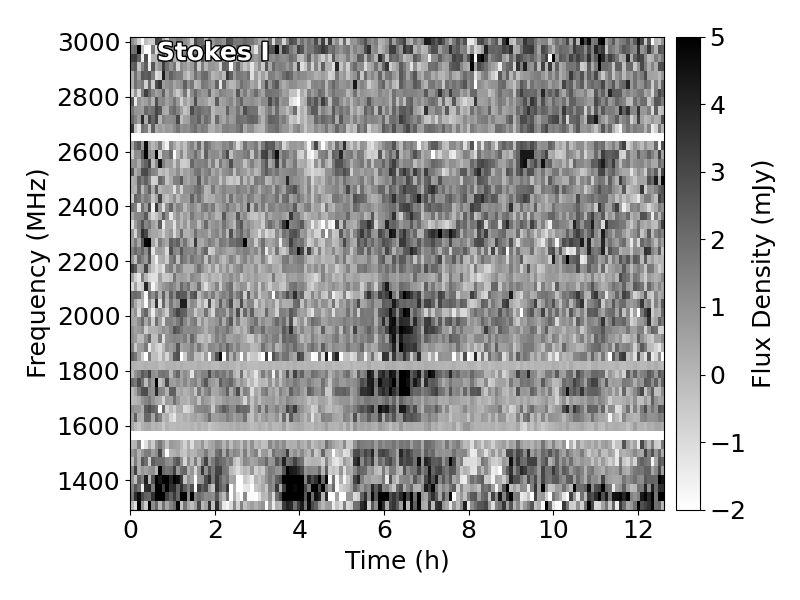}
    \includegraphics[width=\columnwidth]{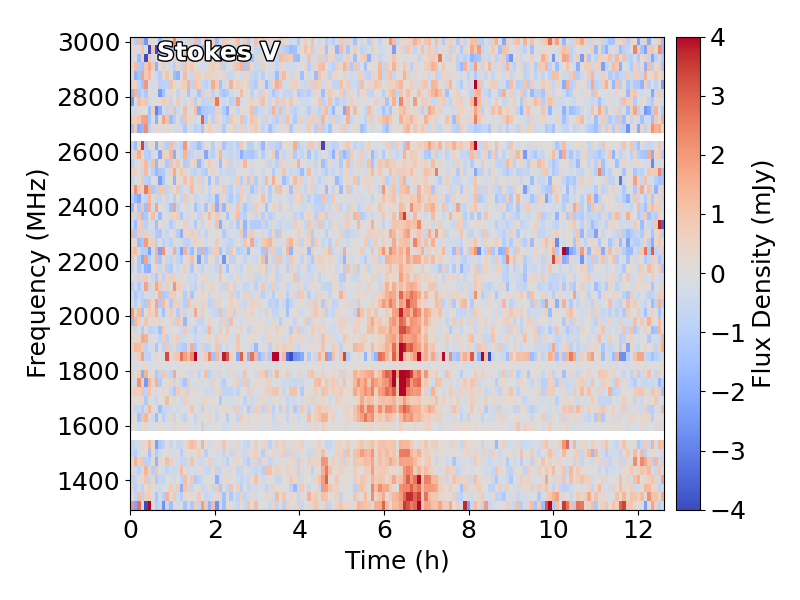}    
    \caption{The dynamic spectra of HD\,105382 in Stokes I and V over $1$--$3$\,GHz, averaged with $5$ minute time resolution and $32$\,MHz frequency resolution.}
    \label{fig:hd105382_IV_DS}
\end{figure}

\begin{figure}
    \centering
    \textbf{\hspace{0.8cm}HD\,105382}\par\medskip
    \includegraphics[trim={0 0 0 0.8cm},clip,width=0.99\textwidth]{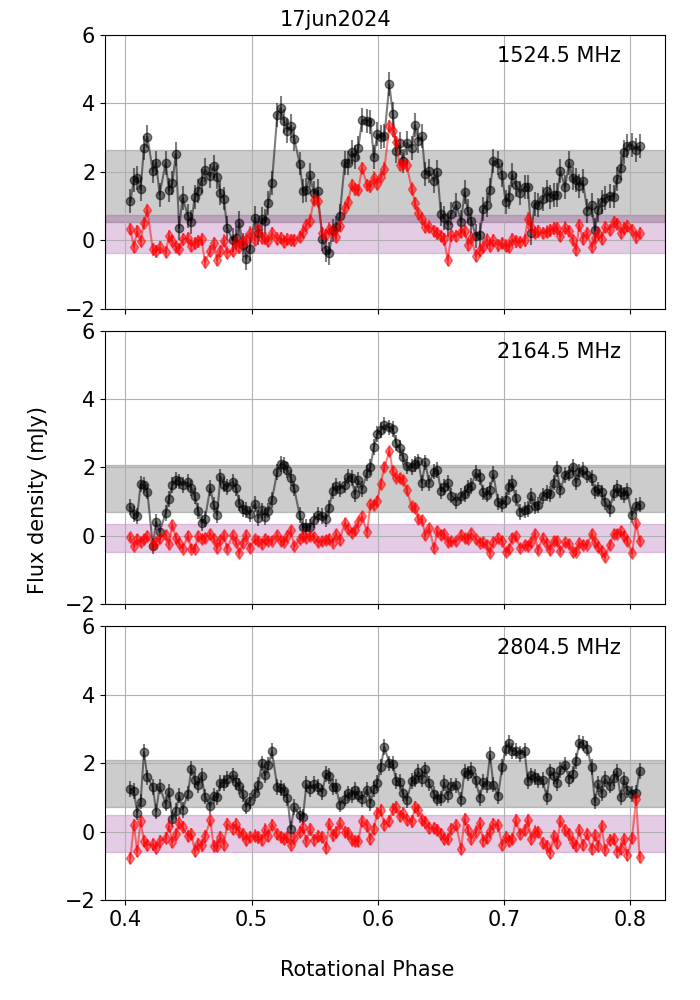}
    \caption{The light curves of HD\,105382 over $1$--$3$\,GHz in Stokes I (black) and Stokes V (red). The integration time for each data point is $5$ minutes. The shaded regions indicate the $3\sigma$ variation about the median flux density away from the phases of enhancement.}
    \label{fig:hd105382_IV_lc}
\end{figure}

HD\,105382 was observed on 2024--06--17. The dynamic spectra (both Stokes I and V) show a very clear enhancement in flux density occurring around the middle of our observations (Figure \ref{fig:hd105382_IV_DS}). The enhancements can also be seen clearly in the light curves obtained within the three sub-bands (Figure \ref{fig:hd105382_IV_lc}). Except for the highest frequency band, the Stokes I light curve exhibits additional enhancements, which can also be seen from the dynamic spectrum in Figure \ref{fig:hd105382_IV_DS}, in addition to the primary enhancement. However, only the primary enhancement shows very high circular polarisation. The rotational phase of arrival of the primary component is very close to that obtained for the brightest enhancement in ASKAP data (Figure \ref{fig:askap_lc}). This confirms that the star produces periodic radio pulses and is an MRP.

\begin{figure}
    \centering
    \includegraphics[width=0.99\textwidth]{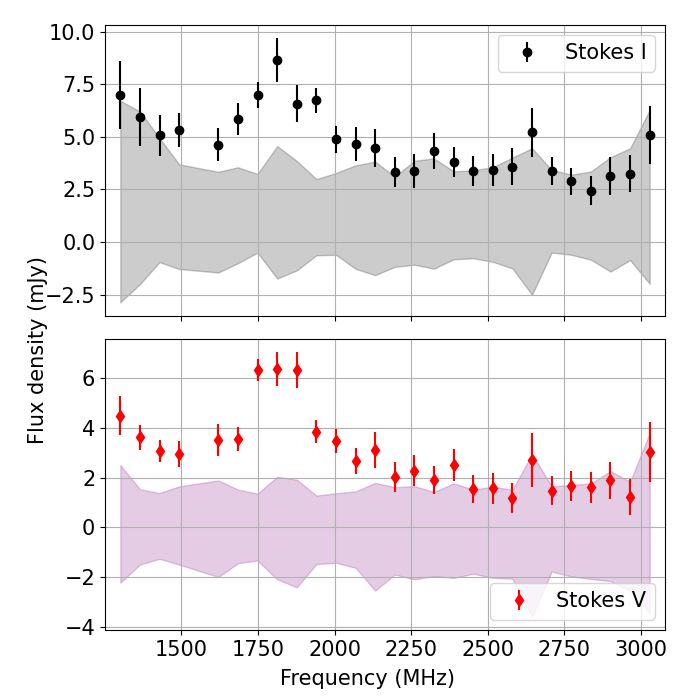}
    \caption{The peak flux density spectrum for the pulse observed from HD\,105382. The shaded regions indicate the basal flux densities $\pm 3\sigma$.}
    \label{fig:hd105382_IV_spectrum}
\end{figure}

The Stokes V dynamic spectrum also suggests that the primary enhancement observed over the phase range $0.1-0.2$ is actually composed of two different components, one of which exhibits a cut-off between 1500 and 1700 MHz and the other cuts off above $2$\,GHz. The cut-off frequencies can be seen more clearly in the peak flux density spectra for the primary pulse, shown in Figure \ref{fig:hd105382_IV_spectrum}. Both Stokes I and V spectra have more than one peak. Such a feature was also reported for ECME from HD\,142990 and attributed to the existence of multiple emission components \citep{das2023}. Based on the Stokes V spectrum (for which the noise in the basal flux density spectrum is lower), the upper cut-off frequency is around 2200 MHz.
The maximum circular polarisation is $96\pm17\%$, observed at $1876.5$\,MHz.

The polar magnetic field strength of the star is $2.6\pm 0.1$ kG \citep{shultz2019c}, corresponding to an electron gyrofrequency of $7.3$\,GHz. Thus, like HD\,83625, this star also exhibits a premature ECME upper cut-off.

\subsection{HD\,149764}\label{subsec:hd149764}
\begin{figure}
    \centering
    \includegraphics[width=0.99\textwidth]{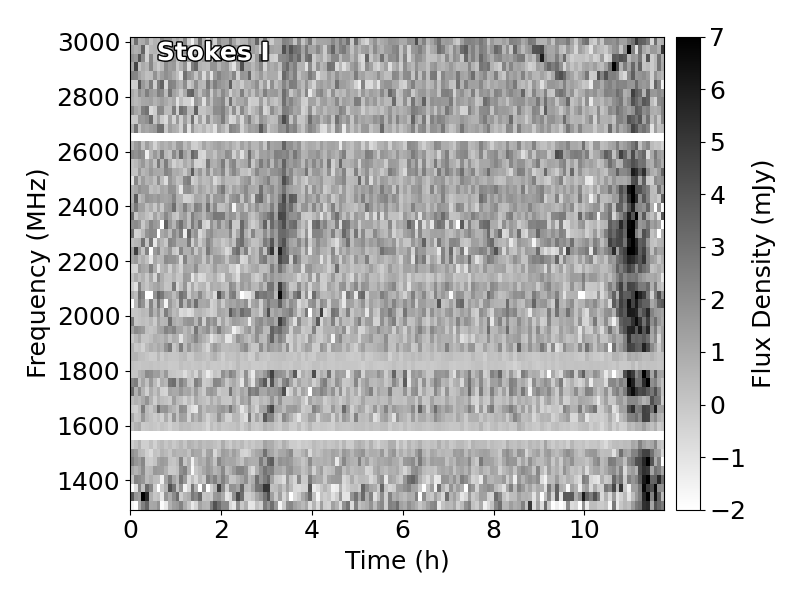}
    \includegraphics[width=0.99\textwidth]{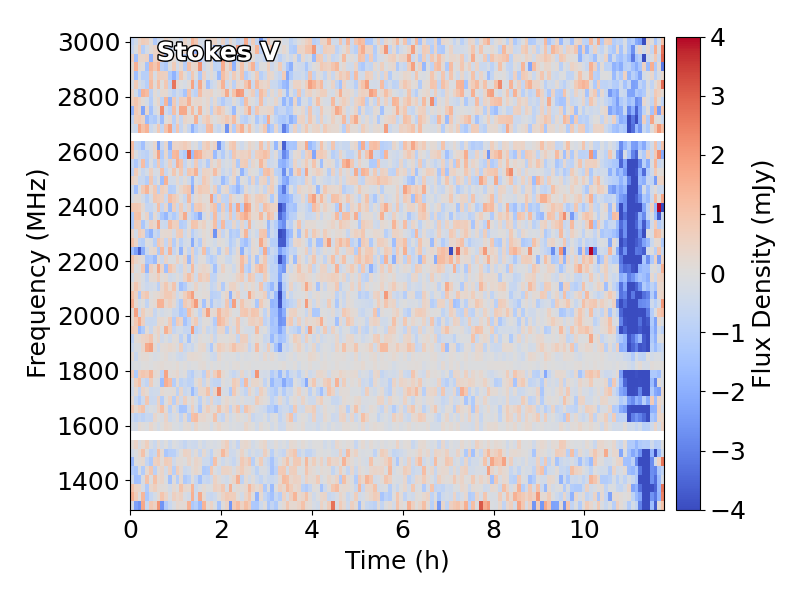}    
    \caption{The dynamic spectra of HD\,149764 in Stokes I and V over $1$--$3$\,GHz, 
    averaged with $5$ minute time resolution and $32$\,MHz frequency resolution. 
    The curve patterns in the top panel (Stokes I) are due to contamination from other sources in the field (rather than emission from our target). Since these emissions are not circularly polarized, the bottom panel (Stokes V) remains free of contamination, clearly highlighting pulses from HD\,149764.}
    \label{fig:hd149764_IV_DS}
\end{figure}

\begin{figure}
    \centering
    \textbf{\hspace{0.8cm}HD\,149764}\par\medskip
    \includegraphics[trim={0 0 0 0.8cm},clip,width=0.99\textwidth]{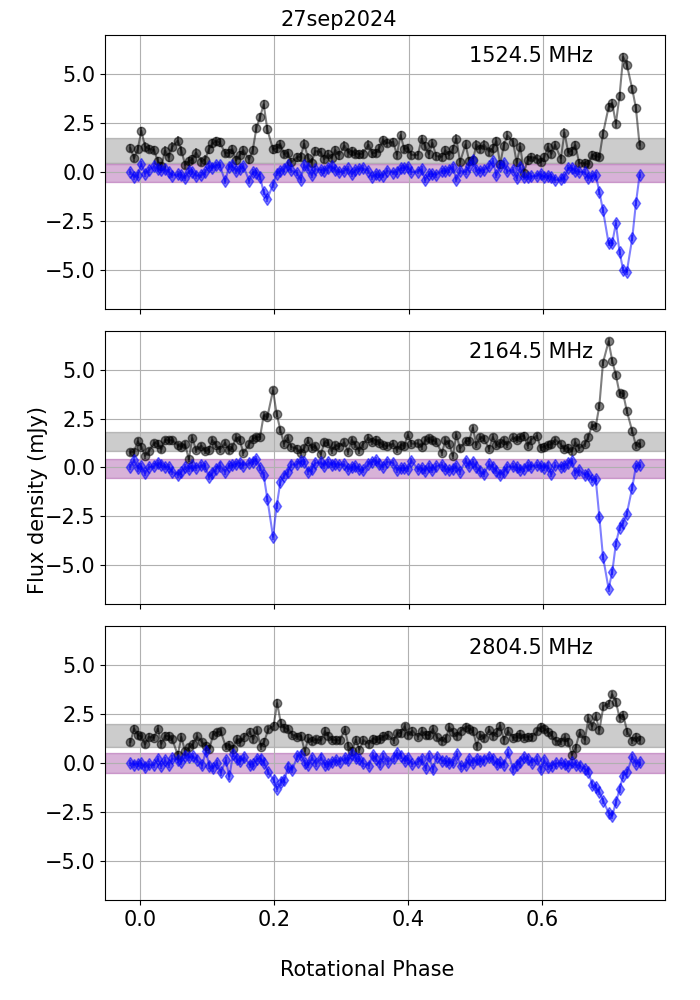}
    \caption{The light curves of HD\,149764 over $1$--$3$\,GHz in Stokes I (black) and Stokes V (blue). The integration time for each point is $5$ minutes. The shaded regions indicate the $3\sigma$ variation about the median flux density away from the phases of enhancement.}
    \label{fig:hd149764_IV_lc}
\end{figure}

\begin{figure*}
    \centering
    \includegraphics[width=0.45\textwidth]{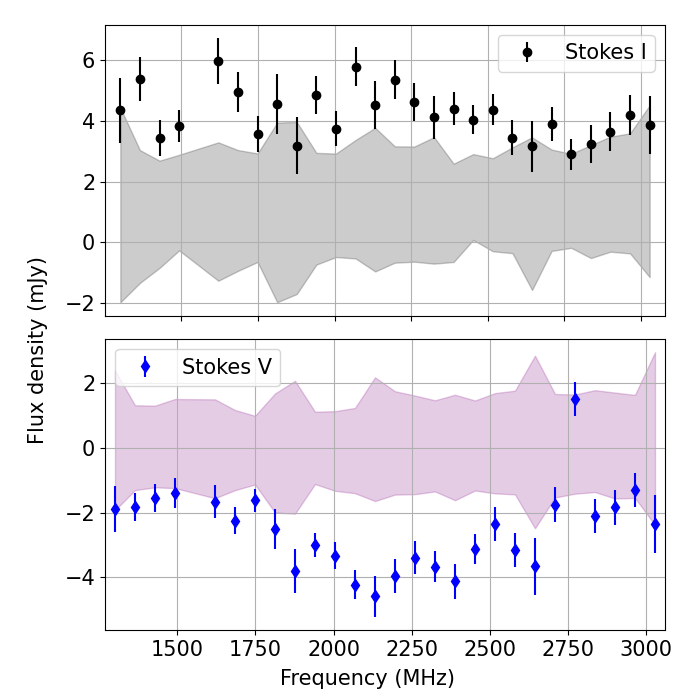}
    \includegraphics[width=0.45\textwidth]{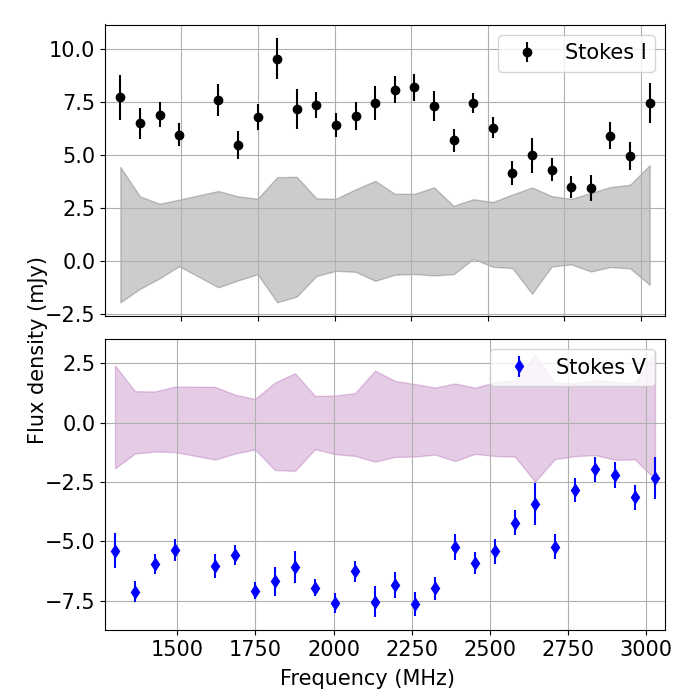}
    \caption{The peak flux density spectrum for the pulse observed from HD\,149764. The shaded regions indicate the basal flux densities $\pm 3\sigma$.}
    \label{fig:hd149764_IV_spectrum}
\end{figure*}

We observed HD\,149764 on 2024--09--27. This is the only star in our campaign that was observed for more than 0.4 rotation cycles. As indicated by the ASKAP light curves, we indeed find that the star exhibits two pulses both of which are left circularly polarised (Figures \ref{fig:hd149764_IV_DS} and \ref{fig:hd149764_IV_lc}). This confirms that the star is an MRP. The pulses sweep in opposite directions on the frequency-time plane (Figure \ref{fig:hd149764_IV_DS}), which is expected to arise from geometric effects \citep{trigilio2011,leto2016,das2020a}. In Figure \ref{fig:hd149764_IV_spectrum}, the peak flux density spectra of both stars are shown (for Stokes V, absolute flux densities are plotted). The leading pulse is the most prominent between 1750 and 2750 MHz (can also be seen from the dynamic spectra in Figure \ref{fig:hd149764_IV_DS}), and reaches 100\% circular polarisation at $2132.5\pm 32$\,MHz. Its upper cut-off frequency is around $2644.5$\,MHz based on both Stokes I and V spectra. The trailing pulse is $\approx 100\%$ circularly polarised below around $2800$\,MHz. The spectra remains flat up to $2260.5$\,MHz and after that, it declines steeply. Following our definition in \S\ref{subsec:hd83625}, we estimate the upper cut-off frequency to be $2836.5\pm 32$ MHz (Stokes I spectrum). This is smaller than the electron gyrofrequency corresponding to the largest of the two \bz~measurements \citep[in absolute values,][]{bagnulo2015}. The lower limit to the star's dipolar magnetic field is $2.9$\,kG \citep{preston1967,bagnulo2015}, which translates to an electron gyrofrequency of $8$\,GHz. Thus, similar to the other two newly discovered MRPs, the ECME of HD\,149764 exhibits a premature upper cutoff.

\subsection{HD\,151965}\label{subsec:hd151965}
\begin{figure}
    \centering
    \includegraphics[width=0.95\textwidth]{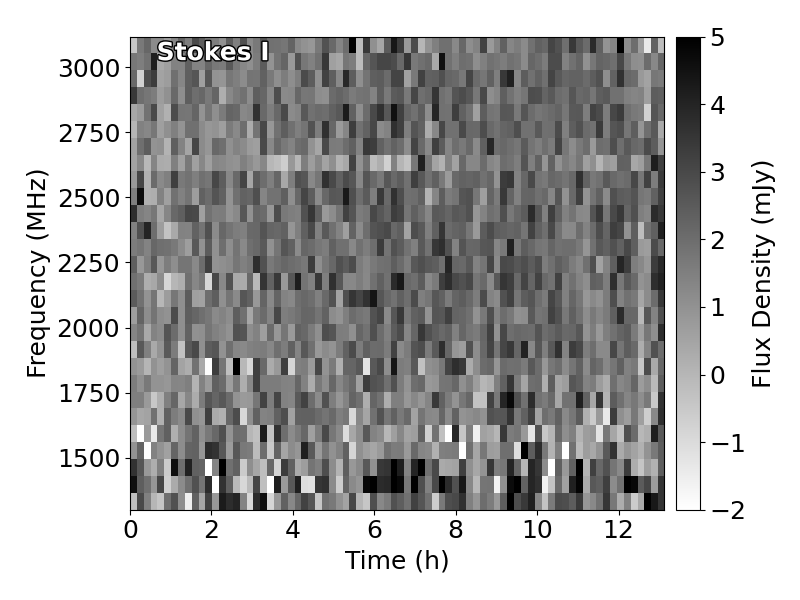}
    \includegraphics[width=0.95\textwidth]{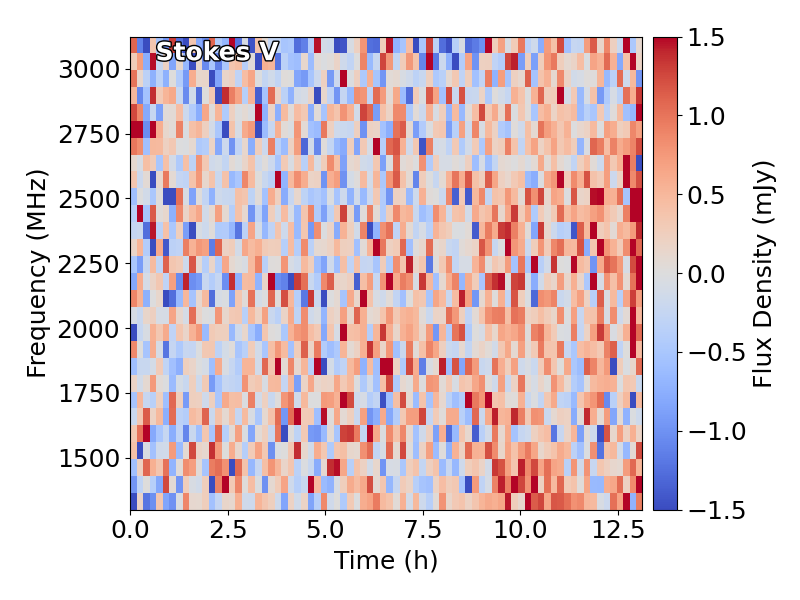}
    \caption{The dynamic spectra for the ATCA data obtained for HD\,151965 in Stokes I (top) and Stokes V (bottom) obtained by averaging the data to a resolution of $10$ minutes and $64$\,MHz, respectively in time and frequency. The times on the horizontal axis are relative times since the start of the observation.}
    \label{fig:hd151965_DS}
\end{figure}

\begin{figure}
    \centering
    \includegraphics[width=0.98\textwidth]{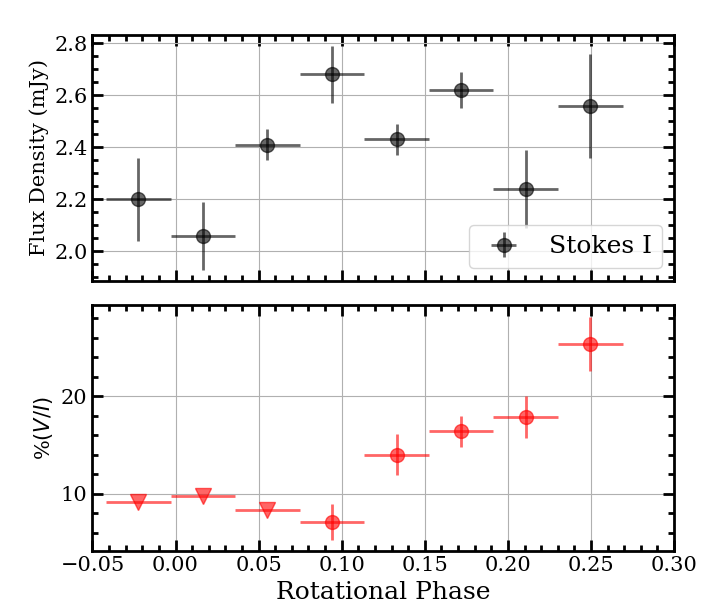}
    \caption{\textbf{Top:} The Stokes I light curve of HD\,151965 obtained from image domain by averaging over the entire available bandwidth. The horizontal error bars represent the averaging time intervals for the respective flux density measurements (1.5 hours or $\approx 0.04$ stellar rotational phase).
    \textbf{Bottom:} The corresponding variation in the fractional circular polarisation.
    }
    \label{fig:hd151965_lc}
\end{figure}

HD\,151965 was observed on 2024--05--10.
The dynamic spectra for Stokes I and V are shown in Figure \ref{fig:hd151965_DS}. We do not find any clear enhancement in flux density in either total intensity or circular polarisation. This motivated us to perform an image-based search for flux density enhancement. The light curve (averaged over the entire frequency range) in total intensity is shown in the top panel of Figure \ref{fig:hd151965_lc} and teh corresponding variation in circular polarisation fraction is shown in the bottom panel. 
The total intensity exhibits hints of a slow rise in flux density during the course of our observation. Such a variation is indicative of incoherent nature of the emission. The circular polarisation, on the other hand, exhibits a very clear increase between 0.2 and 0.25 phases. Unfortunately, the timescale of this variation ($\approx 0.15$ phase) lies in the borderline of the minimum flux density gradient condition so that it does not allow us to conclusively identify the incoherent or coherent nature of the emission.


\section{Discussion}\label{sec:discussion}
We discuss the observed characteristics of ECME from the newly discovered MRPs and compare them with those of the existing MRP sample in the following subsections.

\subsection{Nature of radio emission from HD\,151965}\label{subsec:hd151965_no_ECME}
HD\,151965 is the only star from which we do not detect any convincing signature of ECME. 
With the current observation, we cannot rule out the possibility of the star being an MRP. In particular, its radio emission is peculiar in terms of the following properties:

\begin{enumerate}
    \item Magnitude of the circular polarisation: We observed $\sim 25\%$ circular polarisation at our central frequency of 2 GHz (bottom panel of Figure \ref{fig:hd151965_lc}). This is a very high circular polarisation fraction for incoherent emission at the frequencies under consideration. Until now, HD\,142184 (also known as HR\,5907) is the only magnetic hot star reported to exhibit $\gtrsim 20\%$ circular polarisation for `apparently' incoherent radio emission \citep{pritchard2021}. However, very recently, \citet{biswas2025} discovered that the star's sub-GHz emission is coherent in nature raising the possibility that the same might be the case for its $\sim 1$ GHz emission.
    \item Sign of the circular polarisation: Gyrosynchrotron emission is usually polarised in the extra-ordinary (X) mode for optically thin case; for the optically thick case, the polarisation is quite low \citep[$\lesssim 20\%$,][]{dulk1985}. In the X-mode, Stokes $V$ is positive when \bz~is positive and vice-versa. In our case, we observed positive, high circular polarisation, even though the star's \bz~exhibits consistently negative sign \citep{bohlender1993}. This is a very rare observation and has been reported only once in the past, for the case of HD\,142184 \citep{leto2018}. For HD\,142184 also, \bz~is negative throughout the stellar rotation cycle. Yet, the star exhibits positive Stokes $V$ below 6 GHz (it is negative above 6 GHz, consistent with expectation). Following the discovery of the star being an MRP by \citet{biswas2025}, a viable explanation is that the lower frequency emission is dominated by coherent emission.
\end{enumerate}
To summarise, the similarities observed between HD\,151965 and HD\,142184 strongly suggest that the observed radio emission from the former has a significant contribution from coherent emission.
It is, however, worth mentioning that HD\,151965 has a much stronger magnetic field \citep[$\approx 25$ kG,][Shultz et al. in prep.]{bohlender1993}, which could contribute to enhancing the magnitude of circular polarisation of gyrosynchrotron emission. In addition, theoretically, it is possible to obtain gyrosynchrotron emission in the ordinary mode for the optically thick case under special circumstances \citep{dulk1985}. Thus, while our observations cannot confirm whether or not HD\,151965 is an MRP, they provide a strong motivation for conducting further observations, especially at lower radio frequencies where the circular polarisation of incoherent emission is expected to be even lower.


\subsection{Sense of circular polarisation of confirmed MRPs}\label{subsec:V_sign}
The ideal signature of ECME from a magnetic hot star with an axi-symmetric dipolar magnetic field is a pair of oppositely circularly polarised pulses around each magnetic null \citep{leto2016}. In Stokes V, this is observed as a reversal in the sense of circular polarisation around the magnetic nulls. In our case, we do not see such a reversal for any of the MRPs. This is very likely related to our observation frequency ($1$--$3$\,GHz). Oppositely circularly polarised pulses are produced at opposite magnetic hemispheres. In the absence of any refraction in the stellar magnetosphere, they arrive at the same time leading to zero or low circular polarisation \citep{trigilio2011}. At lower frequencies (sub-GHz), the deviations suffered by the radiation are higher and hence the oppositely circularly polarised pulses are clearly separated in the light curve. As the frequency increases, the difference in pulse arrival times decreases and eventually become zero. This effect has been observed from the MRPs HD\,133880 \citep{das2020} and HD\,35298 \citep{das2022b}. Thus, beyond a certain frequency, the circular polarisation will either go to zero or will be of the same sense as that for the stronger pulse. This critical frequency is likely dependent on the polar magnetic field strength (that defines the emission sites) as well as the magnetospheric plasma.

For HD\,83625, we find low circular polarisation ($\sim +40\%$) for the pulse even though it is prominent in Stokes I until $\lesssim 2$ GHz. Such a case occurs when the Stokes I pulse consists of two oppositely circularly polarised pulses of comparable strengths. Our observation also suggests that the upper cut-off frequencies are similar for pulses produced in both magnetic hemispheres. In the cases of HD\,105382 and HD\,149764, we observe high circular polarisation, which could occur when one of the pulses is significantly weaker than the companion pulse from the opposite magnetic hemisphere. The observation of near $100$\% circular polarisation from HD\,149764 resembles the case of CU\,Vir that exhibits only right circularly polarised ECME above $1.4$\,GHz, which is the upper cut-off frequency for its left circularly polarised pulse. This upper cut-off is much smaller than the corresponding value for the right circularly polarised pulse, which is 3 GHz \citep{das2021}. 
Thus, HD\,149764 likely has significantly different upper cut-off frequencies for the pulses produced in opposite magnetic hemispheres. Once again, both stars should be observed below $1$\,GHz to confirm whether or not they exhibit pulses of both circular polarisations at sub-GHz frequencies, and thereby test our explanation for the lack of observation of reversal in the sign of Stokes $V$.

\subsection{Scaling relation}\label{subsec:scaling_law}

\begin{table*}
\begin{tabular}{lrllcllcc}
\toprule
\headrow Star & $B^0_\mathrm{max}$ & $R\,(R_\odot)$ & $T_\mathrm{eff}\,(\mathrm{kK})$ &  $S_\nu^\mathrm{max, ECME}$ (mJy) & $S_\nu^\mathrm{GS}$ (mJy)& $D$ (pc) & $\frac{L_\mathrm{ECME}\,(\mathrm{erg\,s^{-1}Hz^{-1}})}{10^{16}}$ & $\frac{L_\mathrm{GS}\,(\mathrm{erg\,s^{-1}Hz^{-1}})}{10^{15}} $\\\\
\midrule 
HD\,83625 & $>4.4$ & $2.0^{+1.2}_{-0.3}$ & $12\pm2$ &  $3.61\pm 0.22$ & $0.34\pm 0.07$ & $178\pm2$ & $1.03\pm0.09$ & $1.0\pm0.2$\\
HD\,105382 & $2.6\pm0.1$ & $3.0\pm0.1$ & $18.0\pm0.5$ &  $7.0\pm0.3$ & $1.33\pm0.10$ & $101\pm3$ & $0.65\pm0.07$ & $1.2\pm0.2$\\
HD\,149764 & $>2.9$ & $1.9^{+1.0}_{-0.4}$ & $13\pm 2$ & $9.86\pm0.28$ & $1.03\pm0.09$ & $182\pm2$ & $3.0\pm0.1$ & $3.2\pm 0.6$\\
\midrule
HD\,151965 & $>9.3$ & $2.3\pm0.4$ & $13.5\pm0.6$ & $-$ & $2.1\pm0.1$ & $182\pm2$ & $-$ & $6.2\pm0.4$ \\
\midrule
\caption{The available stellar parameters for the three MRPs (rows 1--3) along with their radio properties reported in this work (rotation periods are already provided in Table \ref{table:ephemeris}). The flux densities correspond to $1.5 $\, GHz. The distances are obtained from GAIA parallaxes \citep{gaia2016}. The stellar parameters for HD\,83625 were obtained from \citet{bagnulo2015} and Shultz et al. in prep.; the parameters for HD\,105382 were obtained from \citet{alecian2011,briquet2001,shultz2018,shultz2019b}; and those for HD\,149764 were obtained from \citet{bagnulo2015,renson2001}.
The bottom row lists the stellar parameters for the star HD\,151965, which, although could not be confirmed as an MRP, exhibits radio properties uncharacteristic of incoherent radio emission (\S\ref{subsec:hd151965_no_ECME}). The corresponding stellar parameters are obtained from \citet{bohlender1993, netopil2017} and Shultz et al. in prep., and the distance is obtained from GAIA parallax \citep{gaia2016}. The incoherent flux density measurement corresponds to a frequency of 2 GHz.
}
\label{table:sample_properties}
\end{tabular}
\end{table*}

Due to the small sample size, it has not been possible to conclusively identify the stellar parameters that drive coherent radio emission. Using a sample of $14$ MRPs, \citet{das2022} first attempted to obtain a scaling relation between spectral radio luminosity at $700$\,MHz and stellar parameters.
The ECME peak spectral luminosity was defined as $L_\mathrm{ECME}=S_\mathrm{peak}d^2$, where $S_\mathrm{peak}$ is the observed peak flux density in either right or left circular polarisation (whichever is the higher) and $d$ is the distance. The stellar parameters considered were
mass, radius, temperature, rotation period and magnetic field strength. They found the strongest correlation with magnetic field strength and stellar effective temperature. The ECME peak luminosity 
was found to exhibit a positive correlation with the polar magnetic field strength. In addition, they found that for stars with effective temperatures below $16.5$ kK, the luminosity increases with temperature, however, beyond that temperature, the luminosity exhibits negative correlation with effective temperature. Based on that, they introduced a quantity called the `$X$-factor', defined as:
\begin{align*}
    X=\frac{B^0_\mathrm{max}}{{(T_\mathrm{eff}-16.5)}^2}
\end{align*}
where $B^0_\mathrm{max}$ is the maximum surface magnetic field strength in kG, and $T_\mathrm{eff}$ is the stellar effective temperature in kK. For dipolar magnetic field, $B^0_\mathrm{max}$ is the polar magnetic field strength at the stellar surface.\footnote{For stars with dipolar magnetic field, the relation between $B^0_\mathrm{max}$ and the longitudinal magnetic field strength \bz~is given by Equation 1 of \citet{preston1967}.}

This relation was investigated in greater detail by \citet{das2022c}, where they also considered the possible correlation between stellar parameters and also relation with incoherent radio emission. Their main conclusions are:
\begin{enumerate}
    \item The relation between $700$\,MHz ECME spectral luminosity and $X$ remains valid after adding three more MRPs to the sample (sample size of $17$). However, for the existing sample, there is a positive correlation between $T_\mathrm{eff}$ and $B^0_\mathrm{max}$ for stars below $T_\mathrm{eff}\leq 16$ kK, thus showing that the positive correlation observed between $L_\mathrm{ECME}$ and $T_\mathrm{eff}$ below $16.5$ kK is unreliable.
    \item For stars with $T_\mathrm{eff}<19$ kK, coherent spectral luminosity positively correlate with the combination of stellar parameters $(B^0_\mathrm{max}R^2)/P_\mathrm{rot}$, which is the governing quantity for incoherent gyrosynchrotron radio luminosity \citep{leto2021,shultz2022,owocki2022}
\end{enumerate}
\citet{das2022} proposed that the correlation with magnetic field strength arises from the dependence of the CBO-driven energy reservoir of the stellar magnetosphere on the magnetic field strength \citep{leto2021,shultz2022,owocki2022}. The dependence on the effective temperature, however, is not understood. Possible reasons include increasing electrons with increasing temperature due to a stronger wind (to explain the increase in the ECME luminosity with temperature), enhanced absorption/inefficient ECME production due to enhanced plasma densities (to explain the decreasing ECME luminosities with temperature beyond a certain value of $T_\mathrm{eff}$), and also simply an artifact of small-sample statistics \citep{das2022,das2022c}. The latter is a prime motivation to expand the sample of MRPs.

In Table \ref{table:sample_properties}, we list the radio and physical properties of the newly discovered MRPs. Among them, HD\,105382 is the only one with an available polar magnetic field strength measurement. For the other two stars (both of which belong to the regime $T_\mathrm{eff}\leq16$ kK), only lower limits to the polar magnetic field strengths are available. As HD\,105382 has an effective temperature of 18 kK, the new MRPs do not help get rid of the correlation between $B_0^\mathrm{max}$ and $T_\mathrm{eff}$. 

We next examine the relation between ECME luminosities and stellar parameters. 
From the MRP sample, we exclude $\rho\,$ Oph C and HR\,5907 as both of them exhibit ECME at all rotational phases making it difficult to disentangle incoherent and coherent components \citep{leto2020b,biswas2025}. Besides, in both cases, \bz~do not exhibit nulls so that the observed flux density is unlikely to be the true estimate of the maximum flux density due to directional effects.

\begin{figure*}
    \centering
    \includegraphics[width=0.99\textwidth]{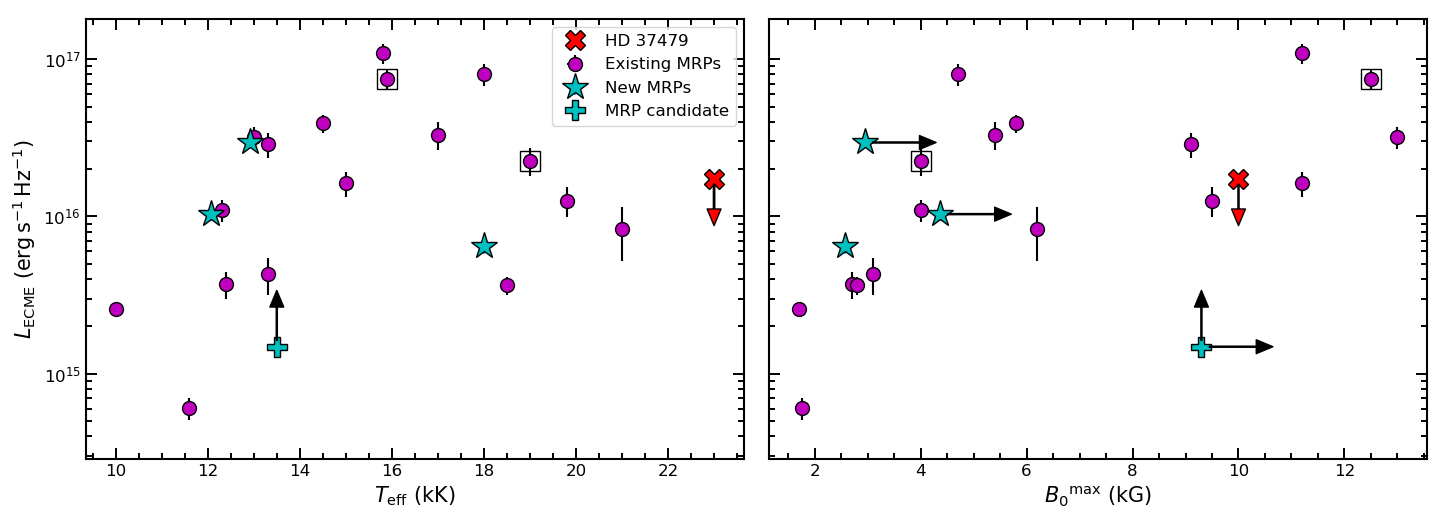}
    \caption{Variation of spectral ECME luminosity with stellar effective temperature (left) and maximum surface magnetic field strength (right). The flux density measurements of all the existing MRPs are obtained at 700 MHz except for those enclosed in squares, for which sub-GHz ECME measurements are not available \citep{leto2019,leto2020}. The flux densities for the newly discovered MRPs, marked as `stars', correspond to a frequency of 1.5 GHz. The only confirmed \textbf{non-}MRP, HD\,37479 is also shown in red. The `$+$' symbol represents HD\,151965, which remains an MRP candidate (see \S\ref{subsec:hd151965_no_ECME}).}
    \label{fig:LECME_B_T}
\end{figure*}

\begin{figure}
    \centering
    \includegraphics[width=0.99\textwidth]{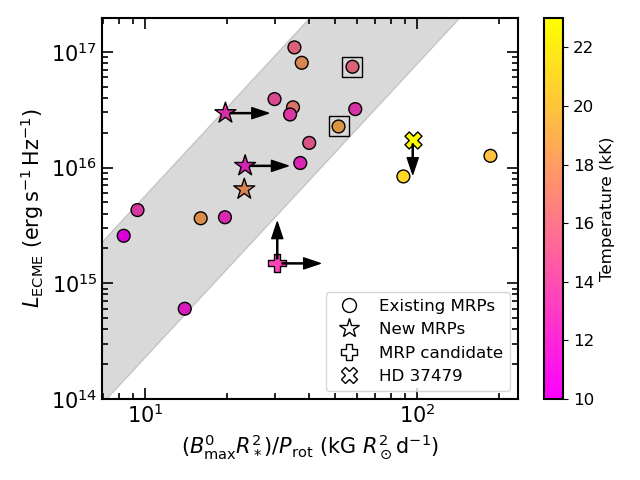}
    \caption{The spectral ECME luminosity against the quantity $(B^0_\mathrm{max}R^2)/P_\mathrm{rot}$, which is the quantity that drives the incoherent radio emission \citep{leto2021,shultz2022,owocki2022}. The new MRPs are highlighted as `stars'. HD\,151965, which remains an MRP candidate, is also shown with a `$+$' symbol.
    }
    \label{fig:LECME_LCBO}
\end{figure}

The majority of the currently known MRPs have flux density measurements at $700$\,MHz. For the three new MRPs, we choose the lowest frequency bin centred at $1.5$\,GHz after dividing the full usable band into three sub-bands. Note that the peak ECME luminosities are obtained after subtracting the basal gyrosynchrotron flux densities from the observed ECME peak flux densities. Figure \ref{fig:LECME_B_T} shows the variation of spectral ECME luminosity $L_\mathrm{ECME}$ with $T_\mathrm{eff}$ and $B^0_\mathrm{max}$. 
Along with the confirmed MRPs, HD\,151965, which could be an MRP (\S\ref{subsec:hd151965_no_ECME}), is also shown with a `+' symbol. Note that for this star, the flux density measurement corresponds to a frequency of 2 GHz. From Figure \ref{fig:hd151965_lc}, assuming that the enhancement arises due to coherent emission (ECME), it is clear that our observations did not cover the full pulse so that the peak flux density is very likely a lower limit to the true peak flux density of the pulse. This aspect is indicated with vertical upward arrows associated with HD\,151965 in Figure \ref{fig:LECME_B_T} and also in Figure \ref{fig:LECME_LCBO}.
We find that the expanded sample still exhibits a parabolic relation with $T_\mathrm{eff}$ and a monotonically increasing relation with $B_0^\mathrm{max}$ as observed by \citet{das2022,das2022c}.


We finally compare $L_\mathrm{ECME}$ with $(B^0_\mathrm{max}R^2)/P_\mathrm{rot}$, a proxy for the centrifugal breakout luminosity \citep{leto2021} and once again, we find that all the newly added MRPs are consistent with the conclusion drawn by \citet{das2022c} (Figure \ref{fig:LECME_LCBO}). Note that the strongest correlation is obtained with magnetic field strength alone, and we still do not have independent evidence for the role of stellar radii or the rotation periods. Excluding the stars with $T_\mathrm{eff}\geq 19$ kK, the Spearman rank correlation co-efficient between $L_\mathrm{ECME}$ and $B_0^\mathrm{max}$ is 0.80 with a p-value of 0.0001 (if we treat the lower limits to magnetic fields as the true values). The correlation deteriorates slightly for $L_\mathrm{ECME}$ and $(B^0_\mathrm{max}R^2)/P_\mathrm{rot}$, with a Spearman rank correlation coefficient of 0.77 and a p-value of 0.0003.
In the future, it will be important to acquire more \bz~measurements for HD\,83625 and HD\,149764 so as to be able to measure their true polar magnetic field strengths, which will be important to use them to derive a robust relation between ECME luminosity and stellar parameters.

\subsection{Future observations}\label{subsec:future}
In the current sample of MRPs, there is a void of stars for which $T_\mathrm{eff}\gtrsim 20$ kK and $B_0^\mathrm{max}\gtrsim 10$ kG. This is an interesting phase-space where HD\,37479 (or $\sigma$ Ori E) lies, which is the only confirmed non-MRP \citep{leto2016,das2022c}. The existing sample appears to indicate that high $T_\mathrm{eff}$ has an adverse effect on ECME production. This is consistent with the lack of ECME from HD\,37479, which has the highest $T_\mathrm{eff}$ among the magnetic hot stars that have searched for ECME. However, this can only be confirmed by examining stars that have strong magnetic fields and high $T_\mathrm{eff}$ so that the effect of the magnetic field and temperature can be disentangled.
Among the newly added MRPs, HD\,105382 is the one with the lowest observed ECME luminosity. However, the star also has the highest $T_\mathrm{eff}$ and the lowest $B_0^\mathrm{max}$ (lower than the available lower limits for the other two) among the three. Thus, it is not possible to uniquely identify the cause behind its low efficiency in ECME production.

There is also a dearth of MRPs with $T_\mathrm{eff}\lesssim 12$ kK and $B_0^\mathrm{max}\gtrsim 10$ kG. Such MRPs will be helpful to confirm or rule out the positive correlation observed between $L_\mathrm{ECME}$ and $T_\mathrm{eff}$ for stars with $T_\mathrm{eff}\lesssim 16$ kK. Thus, future targeted campaigns should focus on well-characterised, strongly magnetic and rapidly rotating stars spanning $T_\mathrm{eff}$ from $10$--$25$ kK. The rapid rotation criterion will help avoid including any variation due to rotation period.

In addition to magnetic field strength and temperature, it is important to investigate the role of stellar rotation period and radius in ECME production, since they have been shown to play a key role for incoherent radio emission. To expand the range of rotation period spanned by the MRP sample, sky surveys offer the most promising route, since targeted campaigns are typically limited by telescope time required. Upcoming facilities such as the DSA2000 \citep{hallinan2024} and the SKA \citep[mid and low,][]{dewdney2009} will play a key role here by rapidly expanding the sample, which will finally help us overcome the issue of small sample size and thus conduct robust statistical tests to understand the ECME production in large-scale stellar magnetospheres.

\section{Conclusion}\label{sec:conclusion}
In this paper, we report the discovery of three new MRPs that were first identified as MRP candidates solely based on the variability in their total intensity light curves constructed from ASKAP survey data. 
This is the first time that the `minimum flux density gradient criterion', introduced by \citet{das2022}, is used to find MRP candidates in all-sky surveys. Among the four candidates obtained this way, we confirm that three of them are MRPs, demonstrating the effectiveness of this strategy. The alternate (and more common strategy) for finding MRP candidates from survey data is to search for highly circularly polarised emission. This strategy, however, cannot find MRPs that do not show high circular polarisations at the relevant frequencies.
For example, the MRP HD\,12447 exhibits negligible circular polarisation for some of its pulses at $700$ MHz \citep{das2022}. Another MRP HD\,142990 exhibits $\lesssim 20\% $ circular polarization at $1.3$ GHz for one of its pulses \citep[][Das et al. under review]{das2023}.
Among the three new MRPs, HD\,83625 exhibits only $\sim 40\%$ circular polarisation when averaged over $64$\,MHz and $5$ minutes. This value will reduce significantly when averaged over longer time and wider bandwidth, making it difficult to detect in conventional circular polarisation surveys even if the observations are acquired around the rotational phase of enhancement.

The `minimum flux density gradient criterion' is a lenient, yet effective strategy that can be employed to find MRP candidates (with known rotation periods) when we have access to multi-epoch data. It can be further generalised as a `variability criterion' where we search for flux density variation by a factor larger than a certain threshold. This criterion is likely to be effective for detecting coherent radio emission from magnetic cool stars as well. Both variability and high circular polarisation criteria have pros and cons, and together they represent a set of highly effective tools to find MRPs (and coherent radio emitters in general) in an unbiased manner.

Although the field of MRPs has seen rapid growth in recent years due to the expansion of the sample, it has not yet been possible to build a robust model of how different stellar parameters govern the phenomenon as the sample is not large enough. As described in the preceding paragraph, all-sky surveys have great potentials to play a key role in overcoming this issue. 
However,
the importance of targeted campaigns will remain, especially to probe regions of stellar parameter phase-space devoid of MRPs (e.g. stars with relatively low effective temperature and strong magnetic field). This is because confirming non-MRPs is equally important as confirming MRPs. Similarly, follow-up wideband observations spanning full rotation cycles will also be important for detailed characterisation of the phenomenon. Only with these extensive observational information, it will be possible to build a complete model for non-thermal radio production in large-scale stellar magnetospheres that successfully explains both incoherent and coherent components.


\newpage 
\begin{acknowledgement}
We thank the referee for their constructive criticism and suggestions.
This scientific work uses data obtained from Inyarrimanha Ilgari Bundara, the CSIRO Murchison Radio-astronomy Observatory. We acknowledge the Wajarri Yamaji People as the Traditional Owners and native title holders of the Observatory site. CSIRO’s ASKAP radio telescope is part of the Australia Telescope National Facility (\url{https://ror.org/05qajvd42}). Operation of ASKAP is funded by the Australian Government with support from the National Collaborative Research Infrastructure Strategy. ASKAP uses the resources of the Pawsey Supercomputing Research Centre. Establishment of ASKAP, Inyarrimanha Ilgari Bundara, the CSIRO Murchison Radio-astronomy Observatory and the Pawsey Supercomputing Research Centre are initiatives of the Australian Government, with support from the Government of Western Australia and the Science and Industry Endowment Fund.
The Australia Telescope Compact Array is part of the Australia Telescope National Facility (\url{https://ror.org/05qajvd42}) which is funded by the Australian Government for operation as a National Facility managed by CSIRO.

KR thanks the LSST-DA Data Science Fellowship Program, which is funded by LSST-DA, the Brinson Foundation, and the Moore Foundation; Their participation in the program has benefited this work.
YW acknowledges support through Australian Research Council Discovery Project DP 220102305.

\end{acknowledgement}

\paragraph{Competing Interests}
None.

\paragraph{Data Availability Statement}
The ASKAP data are available from the CSIRO ASKAP Science Data Archive (CASDA, \url{https://research.csiro.au/casda/}.). The ATCA data and subsequent analysis are available upon request.



\bibliography{hms.bib}

\appendix

\section{Period Determination}\label{sec:rotation_period}

\begin{figure*}
\centering
\begin{tabular}{cc}
    \includegraphics[width=0.5\textwidth]{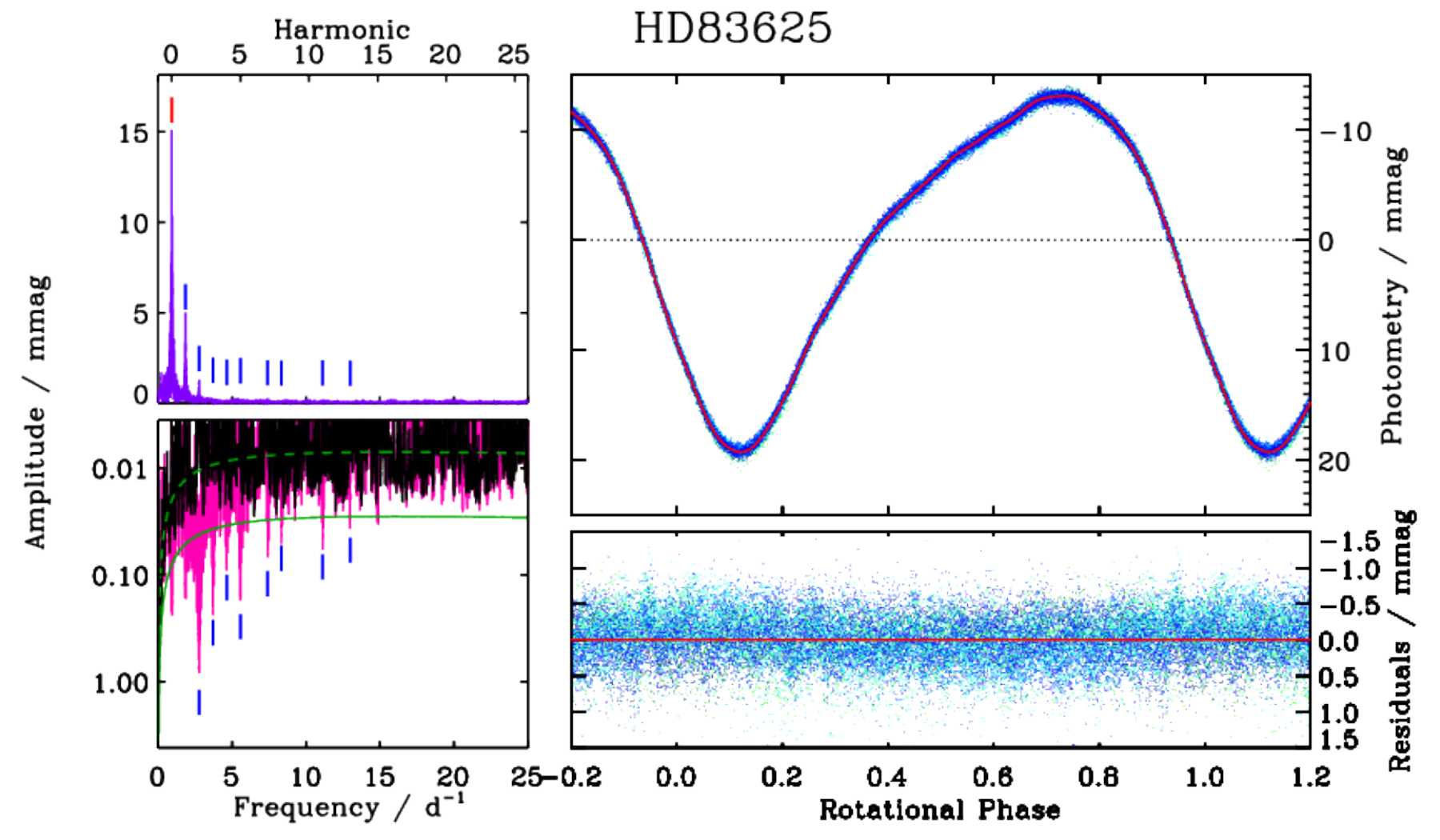} & 
    \includegraphics[width=0.5\textwidth]{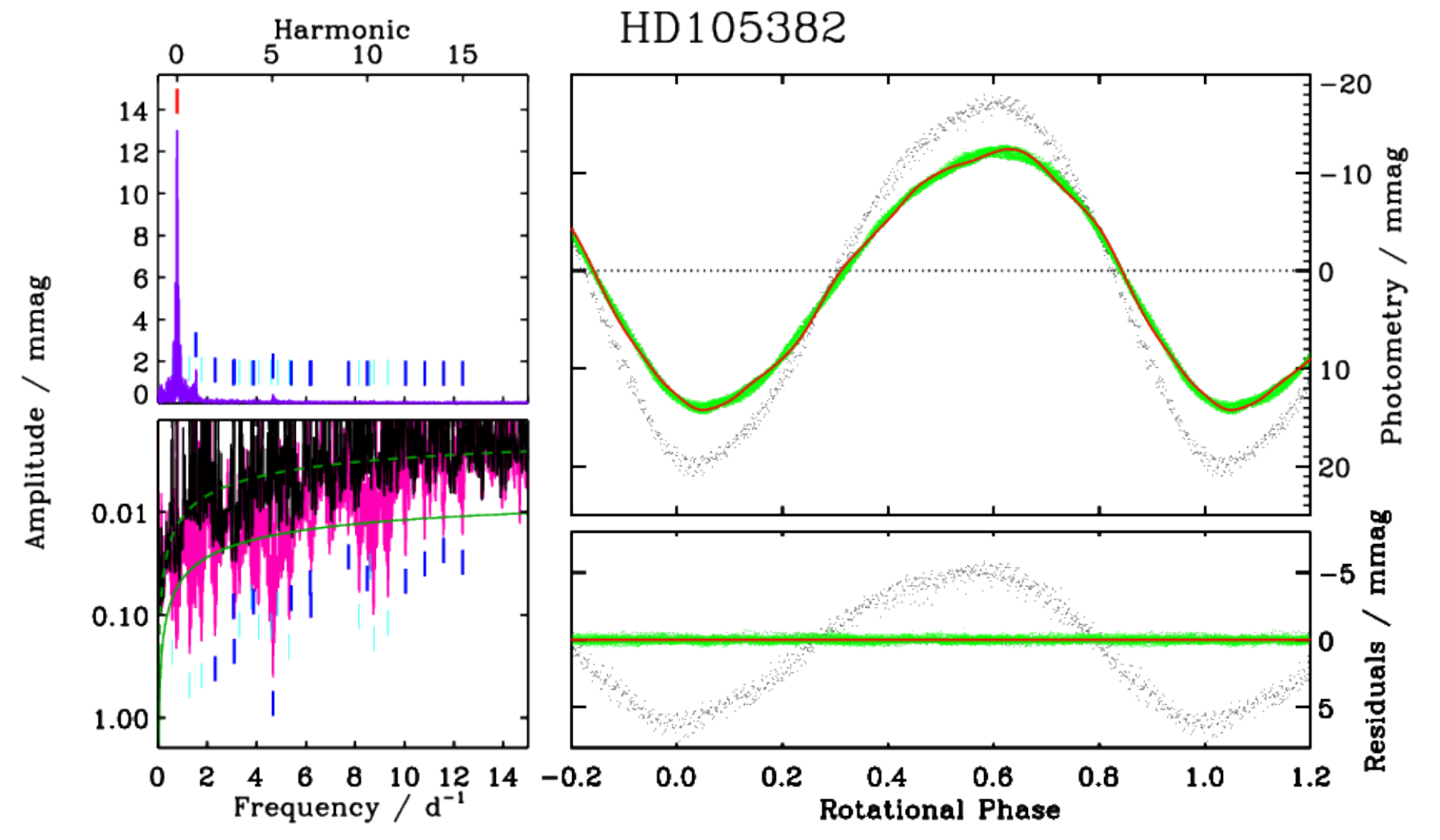} \\
    \includegraphics[width=0.5\textwidth]{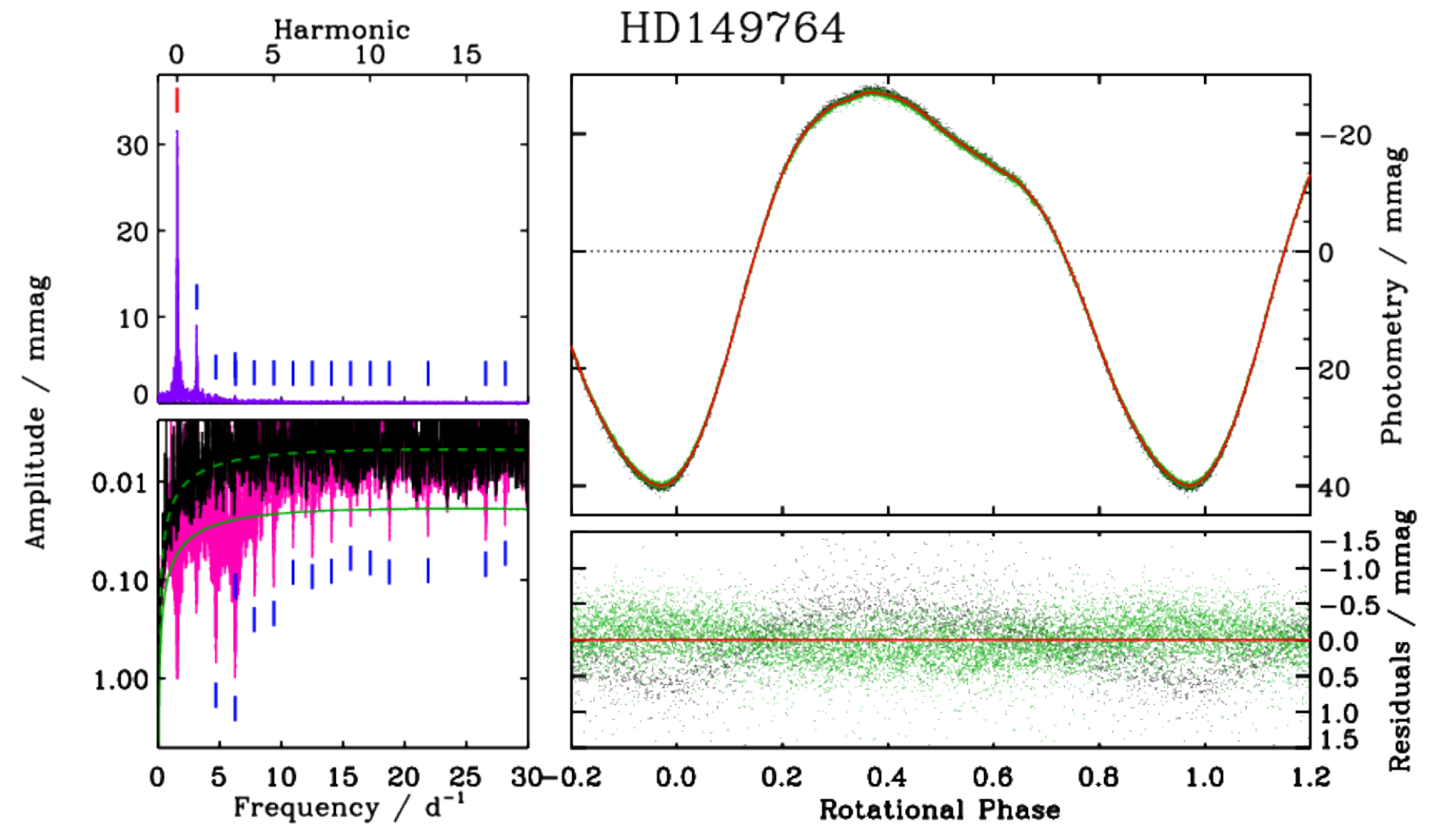} & 
    \includegraphics[width=0.5\textwidth]{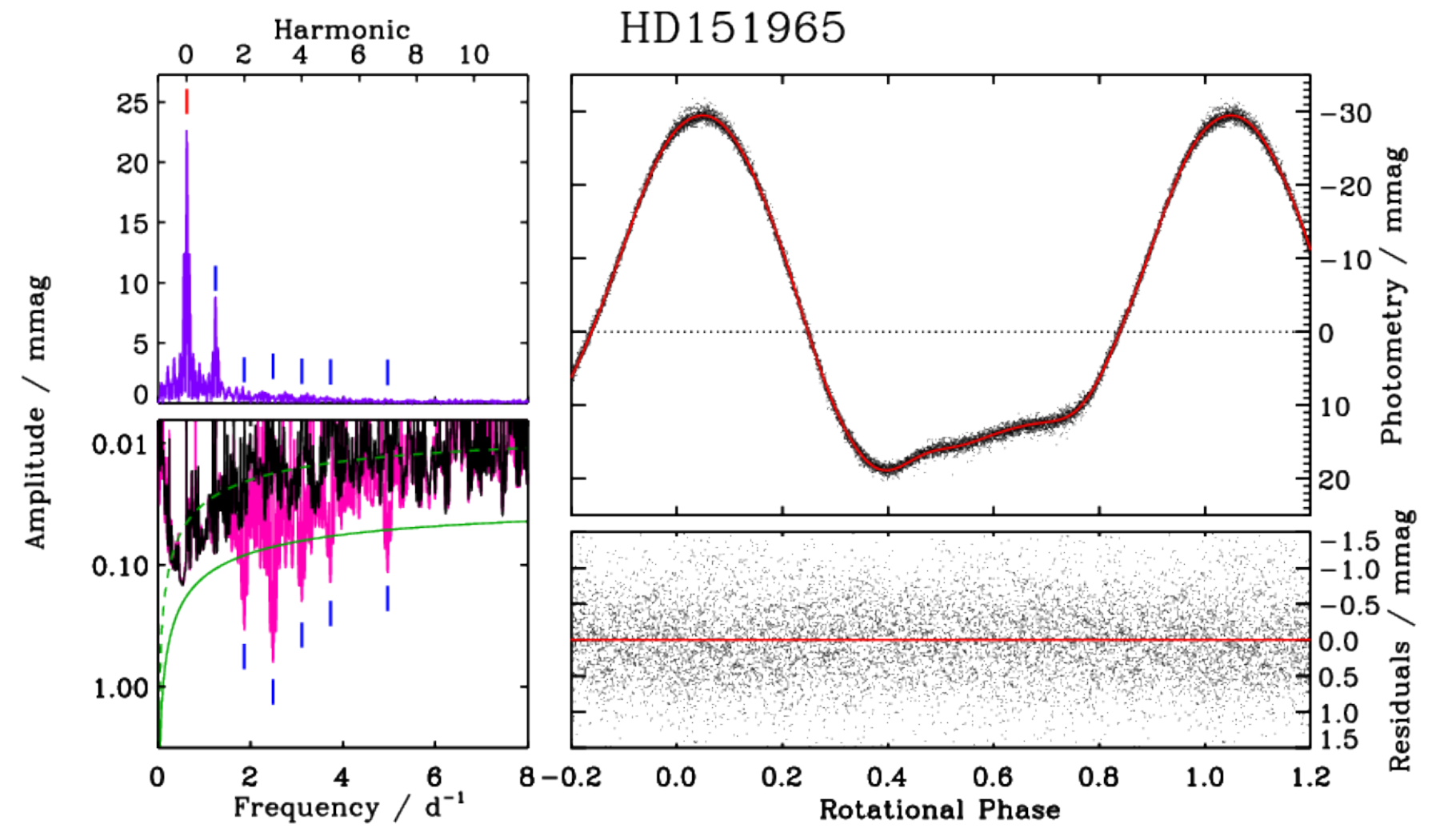} \\
\end{tabular}
    \caption{Period determination from TESS photometry. {\em Left}: full periodogram (top) and (bottom) periodogram after prewhitening with the rotational frequency and its first harmonic (pink) and after prewhitening all significant harmonics of the rotational frequency (black). The dashed green curve shows a noise model fit to the fully prewhitened periodogram; the solid green curve indicates 4$\times$ the noise model, with frequencies above this threshold counting as significant. The red dash indicates the rotational frequency; dark blue dashes indicate harmonics; light blue dashes indicate significant frequencies not associated with rotation. {\em Right}: TESS light curves folded with the rotational frequency (top); different colours indicate different TESS sectors, while the red curve shows the harmonic model. Residuals after subtraction of the model are shown on the bottom.}
    \label{fig:freqs}
\end{figure*}

Periods were determined using TESS light curves. The Transiting Exoplanet Survey Satellite (TESS) is a space telescope obtaining high-precision ($\mu$mag) photometry \citep{2015JATIS...1a4003R}. Its initial mission lasted two years, during which it observed 85\% of the sky in overlapping sectors of $96 \times 24$ deg. Each sector is observed for about 27 days. Data for high-priority targets is downloaded with a 2-minute cadence. The instrument obtains data over a broad bandpass (6000 \AA to 10,000 \AA), with large ($21 \times 21$ arcsecond) pixels. Following completion of the initial two-year run, TESS has continued observing the sky, obtaining data for an extremely large number of stars with varying cadences. 

The light curves for all four stars were downloaded in reduced form from the Mikulski Archive for Space Telescope Data. The light curves are shown in Figure\ \ref{fig:freqs}.

Light curves were analyzed with the {\sc period04} package for Lomb-Scargle frequency analysis \citep{2005CoAst.146...53L}. Periodograms are shown in Figure\ \ref{fig:freqs}. When more than one frequency is present, as is typically the case, the rotational frequency $f_{\rm rot}$ was identified as the first term in a harmonic series. Frequencies were identified via pre-whitening of the light curve with the previously identified frequencies, which {\sc period04} performs via fitting the observations with the function 

\begin{equation}
A(t) = \sum_{i}a_i \sin{(2\pi f_i t + \Phi_i)},
\end{equation}

\noindent where $A(t)$ is the amplitude at time $t$, and $a_i$, $f_i$, and $\Phi_i$ are respectively the amplitude, frequency, and phase offset of the $i^{\rm th}$ term. These harmonic models are shown in Figure\ \ref{fig:freqs}. The light curves were iteratively pre-whitened until no significant frequencies remained, with a significant frequency defined as one for which the signal-to-noise $S/N$ is greater than 4. The frequency-dependent $S/N$ was determined via a low-order polynomial fit to the fully pre-whitened frequency spectrum in log-log space. Harmonics of $f_{\rm rot}$ were identified as those significant frequencies for which the difference between the frequency and an integer multiple of $f_{\rm rot}$ is less than the Rayleigh criterion ($1/\Delta T$ where $\Delta T$ is the timespan of the observations).

Once the final rotation periods were determined, epochs $T_0$ were evaluated based on the time of minimum light. 

\noindent {\bf HD\,83625}: The reported rotation period of the star is $1.07852(5)$ days \citep{heck1987}.
There are 6 sectors of TESS data (9, 10, 36, 37, 62, and 63). The light curve contains 9 harmonics of the rotational frequency, with no additional significant frequencies. The period and epoch are:

$P_{\rm rot} = 1.0784747(1)$~d (consistent with literature value)

$T_0 = 2460015.42769$

\noindent {\bf HD\,105382}: \citet{briquet2001} reported a rotation period of $1.295$ days for this star. 
There are three TESS sectors (10, 37, 64) for this star. The amplitude of the first TESS sector is systematically larger than the other sectors, and this sector was therefore not used in the frequency analysis. In addition to the rotational frequency and its 15 harmonics, the periodogram shows 13 other low-amplitude significant frequencies, possibly associated with pulsation.  The period and epoch are:

$P_{\rm rot} = 1.2950709(2)$~d

$T_0 = 2460041.62838$

\noindent {\bf HD\,149764}: The reported value of the stellar rotation period is $0.63933$ days \citep{renson2001}. For this star,
there are two TESS sectors (39 and 66). The light curve contains 15 harmonics of the rotational frequency, with no additional significant frequencies.  The period and epoch are:

$P_{\rm rot} = 0.63934468(7)$~d (consistent with literature value)\footnote{Although no uncertainty was provided by \citet{renson2001}, the uncertainty has to be $\geq 0.00001$ days, which makes it consistent with the value obtained using TESS data.}

$T_0 = 2460097.82080$

\noindent {\bf HD\,151965}: Using photometric measurements, \citet{lanz1991} reported a rotation period of $1.60841(2)$ days for this star. The star has only one sector (66) of TESS observation. The corresponding light curve contains 6 harmonics of the rotational frequency, with no additional significant frequencies.  The period and epoch are:

$P_{\rm rot} = 1.60866(3)$~d

$T_0 = 2460098.35344$

\noindent Thus the value obtained using TESS data is not consistent with the literature value, which could be related to rotation period evolution \citep[observed for a number of magnetic hot stars, e.g. CU\,Vir, HD\,142990, ][etc.]{mikulasek2011,shultz2019_0}. In this paper, we use the value obtained from TESS data.
\end{document}